\documentclass[reprint,amsmath,amssymb,aps]{revtex4-2}

\makeatletter
\renewcommand{\p@subsection}{}
\renewcommand{\p@subsubsection}{}
\makeatother

\usepackage{changepage}
\usepackage{soul}
\usepackage{graphicx}
\usepackage{dcolumn}
\usepackage{bm}
\usepackage[colorlinks=true, linkcolor=blue, citecolor=red]{hyperref}
\usepackage{floatrow}
\usepackage{mathrsfs}
\newfloatcommand{capbtabbox}{table}[][\FBwidth]

\usepackage{blindtext}
\usepackage[normalem]{ulem}
\usepackage{graphicx}  
\usepackage{dcolumn}   
\usepackage{bm}        
\usepackage{amsmath,amssymb}
\usepackage{hyperref}
\usepackage{color}
\definecolor{purple}{rgb}{0.58,0.0,0.83}
\usepackage[toc,page]{appendix}
\newcommand{\lo}[1]{\textcolor{black}{#1}}

\usepackage[normalem]{ulem}

\begin{document}
\title{Cosmological constraints on the Multi Scalar Field Dark Matter model.}
\author{L. O. T\'ellez-Tovar}
\email{ltellez@fis.cinvestav.mx}
\affiliation{Departamento de F\'isica, Centro de Investigaci\'on y de Estudios Avanzados del IPN, A.P. 14-740, 07000 M\'exico D.F.,
  M\'exico.}
 \affiliation{Instituto de Ciencias F\'isicas, Universidad Nacional Aut\'onoma de M\'exico, Apdo. Postal 48-3, 62251 Cuernavaca, Morelos, M\'exico.}

\author{Tonatiuh Matos}
\affiliation{Departamento de F\'isica, Centro de Investigaci\'on y de Estudios Avanzados del IPN, A.P. 14-740, 07000 M\'exico D.F.,
  M\'exico.}
  \email{tonatiuh.matos@cinvestav.mx}

\author{J. Alberto V\'azquez}
\email{javazquez@icf.unam.mx}
\affiliation{Instituto de Ciencias F\'isicas, Universidad Nacional Aut\'onoma de M\'exico, Apdo. Postal 48-3, 62251 Cuernavaca, Morelos, M\'exico.}

\begin{abstract}
The main aim of this paper is to provide cosmological constraints on the Multi Scalar Field Dark Matter model (MSFDM), in which we assume the dark matter is made up of different ultra-light scalar fields. As a first approximation, we consider they are real and do not interact with each other. We study the equations for both the background and perturbations for $N$-fields and present the evolution of the density parameters, the mass power spectrum and the CMB spectrum. In particular, we focus on two scalar fields with several combinations for the potentials  $V(\phi) = 1/2 m_{\phi}^2 \phi^2$, $V(\phi) = m_{\phi}^2f^2\left[1+\cos(\phi/f)\right]$ and $V(\phi) = m_{\phi}^2f^2\left[\cosh(\phi/f)-1\right]$, however the work, along with the code, could be easily extended to more fields. We use the data from BAO, Big Bang Nucleosynthesis, Lyman-$\alpha$ forest and Supernovae to find constraints on the sampling parameters for the cases of a single field and double field, along with the Bayesian evidence. We found that some combinations of the potentials get penalized through the evidence, however for others there is a preference as good as for the cold dark matter.
 
\end{abstract}

\maketitle
\section{Introduction \label{sec:introd}}

So far the most accepted cosmological model considers the contribution of cold dark matter (CDM) as a key component for structure formation, along with a cosmological constant 
($\Lambda$), as the simplest form of dark energy.
The success of this model, know as $\Lambda$CDM, relies mainly on the accurate agreement with several cosmological observations, for example measurements of the current accelerated expansion of the Universe and the Cosmic Microwave Background Radiation (CMB). 
The best description for the dark matter assumes to be made up of pressureless, non-relativistic, neutral and non-baryonic particles whose interaction is primarily through gravity.
However, the assumption of a particle with these properties brings up many unexplained features, mainly at galactic scales, i.e. the central density behaviour in galactic halos or the overpopulation of substructures at small scales; for an extended review about the problems and possible solutions see Refs. \cite{Weinberg:2013aya, Bull:2015stt}. Recent studies suggest there is no longer a Missing Satellites Problem (MSP), however there could be a problem with so many satellites, see \cite{Kim:2017iwr, DES:2020fxi}.
An alternative that may alleviate these problems is to consider a dark matter but now described by a single scalar field $\phi$ with an associated potential $V\left(\phi\right)$, whose evolution is carried out by the Klein-Gordon equation. The idea of assuming a scalar field as the Dark Matter (DM) of the Universe was introduced about two decades ago, where the simplest possibility is to be real, or complex, minimally coupled to gravity and interacting with ordinary matter only gravitationally \cite{Ji1994,Lee1996,Guzman1997,Guzman2000}. Throughout the years, this model has been rediscovered and received many names, for example: Scalar Field DM (SFDM) \cite{Guzman1997}, fuzzy DM \cite{Hu2000}, Bose-Einstein condensate DM \citep{Boehmer:2007um} and, more recently, ultra-light axion DM \cite{Hui:2016ltb}; here we will refer to it as SFDM,  as it was named in \cite{Matos:1998vk}.  
Based on this idea, the particle associated to the field is an ultralight boson whose mass oscillates around $m_{\phi} \sim 10^{-22}$ eV and hence is able to form Bose-Einstein condensates that conform the galactic structures  \cite{Magana2012, Matos2009, Suarez:2013iw, Urena-Lopez:2019kud, Hui:2016ltb}. In this work we will consider that the scalar field (or fields) are already formed and we will not delve into their origin; see Refs. \cite{Hui:2016ltb, Padilla:2019fju} for details.

For an expanding universe the scalar field cools down along with the expansion, and after a 
while, this causes that the field, the boson gas, freezes and then condensates. 
For an ideal boson gas the condensation temperature goes like $T_c\sim m^{-5/3}$, implying that for a mass big enough the condensation temperature becomes small, but the opposite happens if the mass turns out to be light, or ultra-light, the condensation temperature could be very high. 
However, after the turn around, the \lo{structures} start forming and the re-collapse raises the temperature of the bosons again. Therefore, depending on the initial conditions of the galaxy formation, the boson particles can produce excited states, although most of the boson particles remain in the condensate state, or in the ground state \cite{Matos2009, Magana2012, 2013ApJ...763...19R, Suarez:2013iw, Hernandez-Almada:2017mtm}. These excited particles can be interpreted as other scalar fields. 
Thus, once the galaxy is already formed, if it still contains boson particles in several quantum states, then it can be interpreted as a galaxy with different scalar fields. On the other hand, for heavy particles these vibrations could be neglected, however for ultra-light particles an excitation could be comparable with its original mass. In an effective way the scalar field contains the mass plus the effective mass of the excitation's energy. Thus, this boson gas of particles in excited states could be seen effectively as many scalar fields with different potentials, hence the introduction of the Multi Scalar Field Dark Matter model.

Another motivation to introduce several scalar fields with different potentials could be encouraged because if the 4\% of just the baryonic matter in the universe is so diverse, then we can suspect that the 26\% of the matter, the dark matter, could be made of several species with different properties too. This diversity of particles could be observed and be tested at various scales, for instance at the galactic level by observing the rotation curves as well as at the largest scales of the Universe. If the dark matter is formed of scalar fields, then both results should match flawlessly.  In this work we will focus on the cosmological implications, letting for a parallel work the study of astrophysical features \cite{Atalia2021}.

It has been shown that a scalar field with a convex potential  behaves, in average, like dust during late times and hence mimics the behaviour of the cold dark matter. However, depending on the specific form of the potential and even whether the field is real or complex, it may have different behaviors before acting like a pressure-less fluid. So, in order to have a dark matter evolution, it is necessary that the dependence of the potential with respect to the field is such that it presents a minimum value at some critical point around which the field oscillates \cite{Urena-Lopez:2019kud, Hui:2016ltb, Matos:2008zz}. 
Some examples of such kind of potentials are the parabolic function $V(\phi) = 1/2 m_{\phi}^2 \phi^2$ \cite{Matos2009, Suarez:2013iw, Urena-Lopez:2015gur} and the self-interacting potential with a quartic term contribution $V(\phi) = 1/2 m_{\phi}^2 \phi^2 + \lambda_{\phi} \phi^4$ \cite{Padilla:2019fju, Li:2013nal, Suarez:2015fga, Suarez:2016eez}, or the axion like potential $V(\phi) = m_{\phi}^2f^2\left[1+\cos(\phi/f)\right]$ \cite{Cedeno:2017sou, LinaresCedeno:2020dte, Ross:2016hyb} and its analog $V(\phi) = m_{\phi}^2f^2\left[\cosh(\phi/f)-1\right]$ \cite{Matos2009DynamicsOS, Urena-Lopez:2019xri}. Here  $m_{\phi}$ is interpreted as the mass of the field, $\lambda_{\phi}$ is the self-interacting constant and $f$ represents a decaying constant. 
For a single field, several constrictions on its mass have been imposed by using CMB and matter power spectrum \cite{Hlozek:2014lca}, galactic dynamics \cite{Paredes:2015wga}, dwarf galaxies \cite{Gonzales-Morales:2016mkl, Lora:2011yc, Calabrese:2016hmp}, $N$-body simulations with reionization process \cite{sarkar2016} and Lyman-$\alpha$ flux spectra \cite{Armengaud:2017nkf, Irsic:2017yje}. However, the presence of small inconsistencies among datasets are also found, and can be seen in Figure 1 of Ref. \cite{Padilla:2019fju}, and in Ref. \cite{Marsh:2015xka}. 
This single field model provides a very good description of the evolution of the cosmological densities and the peaks of the CMB as well as the number of substructures in galaxy arrays, among others \cite{Magana2012, Matos2009, Suarez:2013iw, Urena-Lopez:2019kud, Hui:2016ltb}. Nonetheless  it  still presents some open issues \cite{Hui:2016ltb}. For example, numerical simulations have shown that the mass of the field could vary for different scales of the simulation in order to fit the observations, for instance on the formation of galaxies \cite{Mocz:2019pyf}, the whirling plane of satellite galaxies around the Milky Way, Andromeda and Centaurus A galaxies \cite{Muller:2018hks}, the same mass scale in satellite galaxies of the Milky Way \cite{Strigari:2008ib}, or the $\sigma_8$ and $H_0$ tension \cite{Allali:2021azp, Blum:2021oxj}, just to mention a few.

In this work, to alleviate these discrepancies, we open up the possibility that the dark matter may be composed of several types of scalar fields. That is, the main aim of this work is to present a model where the dark matter may be made up by several scalar fields, with different  potentials, and to show its constraints imposed by current cosmological observations.
This model may help us to alleviate the inconsistencies among the constrictions of the mass values obtained by different observations, arguing that they could be different dark matter particles \cite{Arvanitaki:2009fg}. Also, if we consider two scalar fields with different masses, the same mass scale in the satellite galaxies of the Milky Way could be explained, i.e. one type of particle could form the host galaxy and the other the satellites  \cite{Broadhurst:2018fei}. We will refer to this model as the Multi Scalar Field Dark Matter (MSFDM).
Other areas have included similar ideas where two or more  fields are used, for instance a combination of the inflaton and the SFDM \cite{Padilla:2019fju}, two scalar fields as dark energy \cite{Paliathanasis:2014yfa, Vazquez:2020ani}, the inflaton and the curvaton \cite{Benisty:2018fja}, two scalar fields for inflation \cite{Bamba:2015uxa, Vazquez:2018qdg}, interactions between dark energy and dark matter \cite{Bertolami:2012xn} or the axiverse model \cite{Arvanitaki:2009fg, Mehta:2021pwf, Cicoli:2021gss}  (see also \cite{Gutierrez-Luna:2021tmq}).

Given the motivations above, in this paper we study the background dynamics and the linear perturbations of the model. As a first approximation we consider the scalar fields are spatially homogeneous, real and with no interaction among each other, however this part could be easily extended in future works. The paper is organized as follows: In sections \ref{sec:back-dyn} and \ref{Perturbations} we present the equations for the background and linear perturbations. In section \ref{numerical} the evolution is obtained with a modified version of the CLASS code for the background, mass power spectrum and CMB power spectrum for different combinations of potentials. In section \ref{Constraints} we show the model constraints obtained with a modified version of Monte Python code, and finally in the last section we present our conclusions.

\section{Mathematical Background}
\label{Background}

\subsection{Background dynamics}
\label{sec:back-dyn}

Throughout this paper we base our analysis on a flat Universe filled up with the standard components: baryons, dark energy in the form of a cosmological constant ($\Lambda$), photons and neutrinos as relativistic species and dark matter (DM). For the neutrinos, we consider the base model used in \cite{Aghanim:2018eyx}, in which they assumed a normal mass hierarchy \cite{NOvA:2017abs, deSalas:2017kay, Super-Kamiokande:2017yvm, Capozzi:2018ubv}, two massless neutrinos and a massive one with the minimal mass $\sum m_{\nu} = 0.06$ eV. In particular, we assume the DM is described by multiple real scalar fields $\phi_i$ endowed with their corresponding potentials $V_i(\phi_i)$, whereas the rest of the matter components are modeled as perfect fluids. Assuming a Friedmann-Lemaitre-Robertson-Walker metric, the equations of motion for the background dynamics are
\begin{subequations}
\begin{eqnarray}
    \label{Hubbleeq}
    H^2 &=& \frac{\kappa^2}{2} \left( \sum_I \rho_I + \sum_i \rho_{\phi i} \right) \, , \\
    \label{continuityeq}
    \dot{\rho_I} &=& - 3\frac{\dot{a}}{a}\left(\rho_I + p_I\right) \, , \\
        \label{KG1}
    \ddot{\phi_i} &=& -3H\dot{\phi_i} - \partial_{\phi i} V_i(\phi_i) \, .
\end{eqnarray}
\end{subequations}
Here, dots represent derivatives with respect to the cosmic time $t$, $H$ is the Hubble parameter, $\kappa^2 = 8\pi G$, and $\rho_I$ and $p_I$ are the energy density and pressure of the $I$-th fluid species respectively, whereas for the scalar fields we have the associated density and pressure given by the standard expressions
\begin{equation}
  \rho_{\phi i} = (1/2) \dot{\phi}^2_i + V_i(\phi_i), \qquad p_{\phi i} = (1/2) \dot{\phi}^2_i - V_i(\phi_i).
\end{equation}
Notice that we are assuming different species of scalar fields, represented each one by the subindex $i$ in the above equations. 
The Klein-Gordon equations~\eqref{KG1}, for each of the fields, can be written in a more manageable form by using the following polar transformation~\cite{Urena-Lopez:2015odd}
\begin{subequations}
\begin{equation}
    \frac{\kappa \dot{\phi_i}}{\sqrt{6}H} \equiv \Omega_{\phi i}^{1/2} \sin (\theta_i/2) \, , \quad
    \frac{\kappa V_i^{1/2}}{\sqrt{3}H} \equiv \Omega_{\phi i}^{1/2}\cos ( \theta_i/2 ) \, , 
\end{equation}
where $\Omega_{\phi i} \equiv \kappa^2 \rho_{\phi i}/3H^2$ represents the dimensionless density parameter, and similarly $\theta_i$ is an angular degree of freedom directly related to the equation of state (EoS) for each one of the fields, $w_{\phi i} \equiv p_{\phi i} /\rho_{\phi i} = -\cos \theta_i$. Additionally, we define the potential variables $y_{1i}$ and $y_{2i}$ as
\begin{equation}
    y_{1i} \equiv - 2\sqrt{2} \frac{\partial_{\phi i} V_i^{1/2}}{H} \, , \quad y_{2i} \equiv - 4 \sqrt{3}  \frac{\partial_{\phi i}^2 V_i^{1/2}}{\kappa H} \, , \label{eq:yvars}
\end{equation}
\end{subequations}
whose form depend on the potential for a particular field. The equivalence between the polar transformation and the fluid equations can be seen in \cite{Cookmeyer:2019rna, Passaglia:2022bcr}.

As a proof of the concept, we focus our study on the following potentials 
\begin{equation}
    V_i(\phi_i) = \left\{ 
    \begin{array}{lcl}
        m^2_{\phi i} f^2_i \left[1 + \cos(\phi_i/f_i) \right] & & \mathrm{cos} \\
        (1/2) m^2_{\phi i} \phi^2_i & & \mathrm{quadratic} \\
        m^2_{\phi i} f^2_i \left[\cosh(\phi_i/f_i) -1 \right] & & \mathrm{cosh} 
    \end{array}
    \right. \, . \label{eq:potentials}
\end{equation}
and their possible combinations; with $f_i$ being a characteristic energy scale for the scalar fields $\phi_i$, and $m_{\phi i}$ its corresponding mass scale. It can be seen that the variables~\eqref{eq:yvars} for the aforementioned potentials can be  written as
\begin{subequations}
\begin{eqnarray}
    y^2_{1i} &=& 4 \frac{m^2_{\phi i}}{H^2} - 2 \lambda_{\phi i} \Omega_{\phi i} \, , \label{eq:yvars1a} \\
    y_{2i} &=& \lambda_{\phi i} y_{1 i} \, . \label{eq:yvars1b}
\end{eqnarray}
\end{subequations}
Notice that the three functional forms in~\eqref{eq:potentials} can be compressed into a  dimensionless parameter, $\lambda_{\phi i} = 3/\kappa^2f_i^2$, which facilitates the numerical calculations. Positive values of $\lambda_{\phi i} > 0$ describe the cosine potential and negative ones $\lambda_{\phi i} < 0$ the cosh potential, whereas the quadratic case corresponds to $\lambda_{\phi i} =0$ (for more details see~\cite{Urena-Lopez:2015gur,Cedeno:2017sou,Urena-Lopez:2019xri}).
Then, for each field,  the associated Klein-Gordon equation~\eqref{KG1} is represented by the following set of coupled equations
\begin{subequations}
\label{difpolaresgeneral}
\begin{eqnarray}
    \theta^\prime_i &=& -3\sin \theta_i  + y_{1i} \, , \\
    \Omega^\prime_{\phi i} &=& 3\left( w_{tot} + \cos \theta_i \right) \Omega_{\phi i} \, , \\
    y^\prime_{1i} &=& \frac{3}{2} \left(1 + w_{tot}\right) y_{1i} + \frac{1}{2} \lambda_{\phi i} \Omega_{\phi i}^{1/2} \sin \theta_i \, ,
\end{eqnarray}
\end{subequations}
with $w_{tot} = \sum_I \Omega_I w_I + \sum_i \Omega_i w_i$, where $\Omega_I \equiv \kappa^2 \rho_I/3H^2$ and $w_I = p_I/\rho_I$. The prime denotes derivative with respect to the number of e-folds $N= \ln a$, and for any given variable $q$ we have the relationship $\dot{q} = Hq^\prime$.
\\

The initial conditions necessary to solve these equations can be seen in Eq.~(2.16) of Ref.~\cite{Urena-Lopez:2015gur} for the quadratic case, for the cosine potential see Eq.~(5)  Ref.~\cite{Cedeno:2017sou} and for the hyperbolic cosine see Eq.~(2.6) in Ref.~\cite{Urena-Lopez:2019xri}. The main purpose in all cases is to match a given value of the density parameter $\Omega_{\phi i,0}$ at the present time
 with the initial values of the dynamical quantities $(\theta_i,y_{1i},\Omega_{\phi i})_{\mathrm{ini}}$ at early enough times (typically for a scale factor of the order of $a_{\mathrm{ini}} \simeq 10^{-14}$). 

In general terms, the mass parameter $m_{\phi i}$ determines the start of the rapid oscillations of the field $\phi_i$ around the minimum of the potential $V_i$, which happens at around $H \simeq 3 m_{\phi i}$. For $\lambda_{\phi i} \neq 0$, Eq.~\eqref{eq:yvars1a} becomes a constraint equation that must be satisfied by the field variables at all times, whereas for $\lambda_{\phi i} = 0$ it simply tells us that $y_{1i} = 2m_{\phi i}/H$. In fact, one expects that at late times $m_{\phi i} \gg H$ so that for the three cases in~\eqref{eq:potentials} the relation $y_{1i} = 2m_{\phi i}/H$ should be satisfied with high accuracy.
Only for the case $\lambda_{\phi i} < 0$ (cosh potential) it is also necessary to impose the scaling solution during radiation domination: $\Omega_{\phi i} = -12/\lambda_{\phi i}$ and $\cos \theta_i = -1/3$, and then the initial value of $y_{1i}$ is calculated from Eq.~\eqref{eq:yvars1a}. Furthermore, the field mass $m_{\phi i}$ is not an independent parameter in this case, and it has been shown that the two parameters are related through~\cite{Urena-Lopez:2019xri}
\begin{equation}
    \frac{m_{\phi i}}{H_{\mathrm{ini}}} = 1.5 \left[ \left( \frac{\lambda_{\phi i}}{3} -4 \right) \frac{\Omega_{\phi i,0}}{\Omega_{r,0}} a_{\mathrm{ini}} \right]^2 \, ,
\end{equation}
with $\Omega_{r,0}$ the present density parameter of relativistic species and $a_{\mathrm{ini}}$ the initial value of the scale factor.

\subsection{Linear density perturbations}
\label{Perturbations}

We consider the linear perturbations for the scalar fields by expanding the field to the leading order, with  $\phi_i\left(\Vec{x},t\right) = \phi_i\left(t\right) + \varphi_i\left(\Vec{x},t\right)$, where $\phi_i(t)$ are the background fields described in the above section, whereas $\varphi_i$ are the field linear perturbations. The perturbed metric, in the synchronous gauge is $ds^2 = -dt^2 + a^2(t)\left(\delta_{lm} + h_{lm}\right)dx^ldx^m$, with $h_{lm}$ being the tensor perturbations of the metric. Working in Fourier space, the perturbed Klein-Gordon equation for each field is given by
\begin{equation}
    \label{perturbedKG}
    \ddot{\varphi}_i = -3H\dot{\varphi}_i - \left( \frac{k^2}{a^2} + \partial^2_{\phi i} V_i \right) \varphi_i - \frac{1}{2}\dot{h}\dot{\phi_i} \, .
\end{equation}
In Eq.~\eqref{perturbedKG}, $k$ is the comoving wavenumber, $h$ is the trace of $h_{lm}$ and $\dot{h}$ is known as the metric continuity. Following the idea presented for the background in the previous section, we use the polar variables \cite{Urena-Lopez:2015gur, Cedeno:2017sou},
\begin{eqnarray}
    \label{perturbedchange}
    \sqrt{\frac{2}{3}}\frac{\kappa \dot{\varphi_i}}{H} &=& -\Omega_{\phi i}^{1/2}e^{\alpha_i}\cos\left(\frac{\vartheta_i}{2}\right) \, , \nonumber \\
    \frac{\kappa y_{i,1} \varphi_i}{\sqrt{6}} &=& -\Omega_{\phi i}^{1/2}e^{\alpha_i}\sin\left(\frac{\vartheta_i}{2}\right) \, ,
\end{eqnarray}
where $\alpha_i$ and $\vartheta_i$ are the new perturbation quantities. If we define the new quantities
\begin{eqnarray}
    \label{perturbedpolarchange}
    \delta_{0i} &=& -e^{\alpha_i}\sin\left(\frac{\theta_i - \vartheta_i}{2}\right) \, , \nonumber \\
    \delta_{1i} &=& -e^{\alpha_i}\cos\left(\frac{\theta_i - \vartheta_i}{2}\right) \, ,
\end{eqnarray}
where the density contrast is $\delta_{\phi i} \equiv \delta\rho_{\phi i}/\rho_{\phi i} = \delta_{0i}$, then 
the perturbed Klein-Gordon equation~\eqref{perturbedKG} can be rewritten as
\begin{widetext}
\begin{subequations}
\label{generalperturbedequation}
\begin{eqnarray}
    \label{generalperturbedequation1}
    \delta^\prime_{0i} &=& -\left[3 \sin \theta_i + \frac{k^2}{k^2_{Ji}} ( 1-\cos \theta_i) \right] \delta_{1i} + \frac{k^2}{k^2_{Ji}} \sin\left(\theta_i\right)\delta_{0i} - \frac{1}{2} h^\prime \left(1 - \cos \theta_i \right) \, , \\
    \delta^\prime_{1i} &=& - \left[3 \cos \theta_i + \left( \frac{k^2}{k^2_{Ji}} - \frac{\lambda_{\phi i} \Omega_{\phi i}}{2 y_{1i}} \right) \sin \theta_i \right] \delta_{1i} + \left( \frac{k^2}{k^2_{Ji}} - \frac{\lambda_{\phi i} \Omega_{\phi i}}{2 y_{1i}} \right) \left(1 + \cos \theta_i \right) \delta_{0i} - \frac{1}{2} h^\prime \sin \theta_i \, , \label{generalperturbedequation2} 
\end{eqnarray}
\end{subequations}
\end{widetext}
where we have introduced the Jeans wavenumber as $k^2_{Ji} = H^2 a^2 y_{1i}$. Other quantities of interest are the perturbations for the energy density $\delta\rho_{\phi i}$, pressure $\delta p_{\phi i}$ and velocity divergence $\Theta_{\phi i}$, which are 
\begin{subequations}
\begin{eqnarray}
    \label{perturbedquantities1}
    \delta\rho_{\phi i} &=& \dot{\phi_i} \dot{\varphi_i} + \partial_{\phi i} V \varphi_i \, , \\
    \delta p_{\phi i} &=& \dot{\phi_i} \dot{\varphi_i} - \partial_{\phi i} V \varphi_i \, , \\
    \left(\rho_{\phi i} + p_{\phi i}\right) \Theta_{\phi i} &=& \frac{k^2}{a} \dot{\phi_i} \varphi_i \, . 
\end{eqnarray}
\end{subequations}
In terms of the new variables $\delta_{0i}$ and $\delta_{1i}$, they are written as
\begin{subequations}
\begin{eqnarray}
    \label{perturbedquantities2}
    \delta \rho_{\phi i} &=& \delta_{0i} \, \rho_{\phi i} \, , \\
    \delta p_{\phi i} &=& \left( \delta_{1i} \sin \theta_i - \delta_{0i} \cos \theta_i \right)  \, \rho_{\phi i} \, , \\
    \label{perturbeddivergencevelocity}
    \left(\rho_{\phi i} + p_{\phi i}\right) \Theta_{\phi i} &=& \frac{k^2\rho_{\phi i}}{aHy_{i,1}} \left[ \left(1 - \cos \theta_i \right) \delta_{1i} - \sin \theta_i \delta_{0i} \right] \, . \nonumber \\
\end{eqnarray}
\end{subequations}

Again, depending on the value of $\lambda_{\phi i}$, one recovers the perturbed equations of any of the three different potentials of this work~\eqref{eq:potentials}. The initial conditions for the perturbations simply are $\delta_{0 \mathrm{init}} = 0$ and $\delta_{1 \mathrm{init}} = 0$. It has been shown that the dynamical variables $\delta_{0i}$ and $\delta_{1i}$ quickly reach an attractor behavior driven by the non-homogeneous term $h^\prime$ in Eqs.~\eqref{generalperturbedequation}~\cite{Urena-Lopez:2015gur, Cedeno:2017sou, Urena-Lopez:2019xri}.
%

\subsection{Numerical Results}
\label{numerical}


In this section we show the background evolution, mass power spectrum (MPS) and CMB power spectrum for different combinations of potentials, obtained with a modified version of the CLASS code that is able to deal with multiple scalar fields~\cite{class, Urena-Lopez:2015gur, Cedeno:2017sou}. This version of the code is publicly available and can be found in \cite{two_sfdm_class}. We use the ratio $R = \Omega_{\phi1,0}/\Omega_\text{DM,0}$ to parameterize the energy density of the scalar fields, where $\Omega_\text{DM,0}= \Omega_{\phi1,0}  + \Omega_{\phi2,0} + \Omega_{{\rm cdm},0}$ represents the current total dark matter contribution from the scalar fields sector. The combination of the fields is symmetric, then for reference we take $\phi_1$ to define $R$, otherwise if $\phi_j$ is taken as reference we need to redefine $R = \Omega_{\phi j}/ \left(\sum_i \Omega_{\phi i} + \Omega_{{\rm cdm}}\right)$.
The mass values for the fields were taken from the references mentioned in the introduction, in particular, those reported in \cite{Broadhurst:2018fei}, where the existence of at least two scalar fields with masses of $10^{-22}$ and $10^{-20}$ eV is proposed to explain the observations of the galaxy cores.

Throughout this work we consider several combinations for the fields, and combined them with the CDM as well.
This is because, until now, the nature of the dark matter is not yet determined and several possibilities should bear in mind. As the next approximation to the single field, here we assume two types of fields with different combinations of potentials  as dark matter, however, once we have enough accurate data,  the analysis (and the code) could be easily  generalized to $N$ different fields.
\\

\begin{figure*}
    \centering
    \makebox[14cm][c]{
    \includegraphics[trim = 0mm  0mm 10mm 0mm, clip, width=10.cm, height=7.5cm]{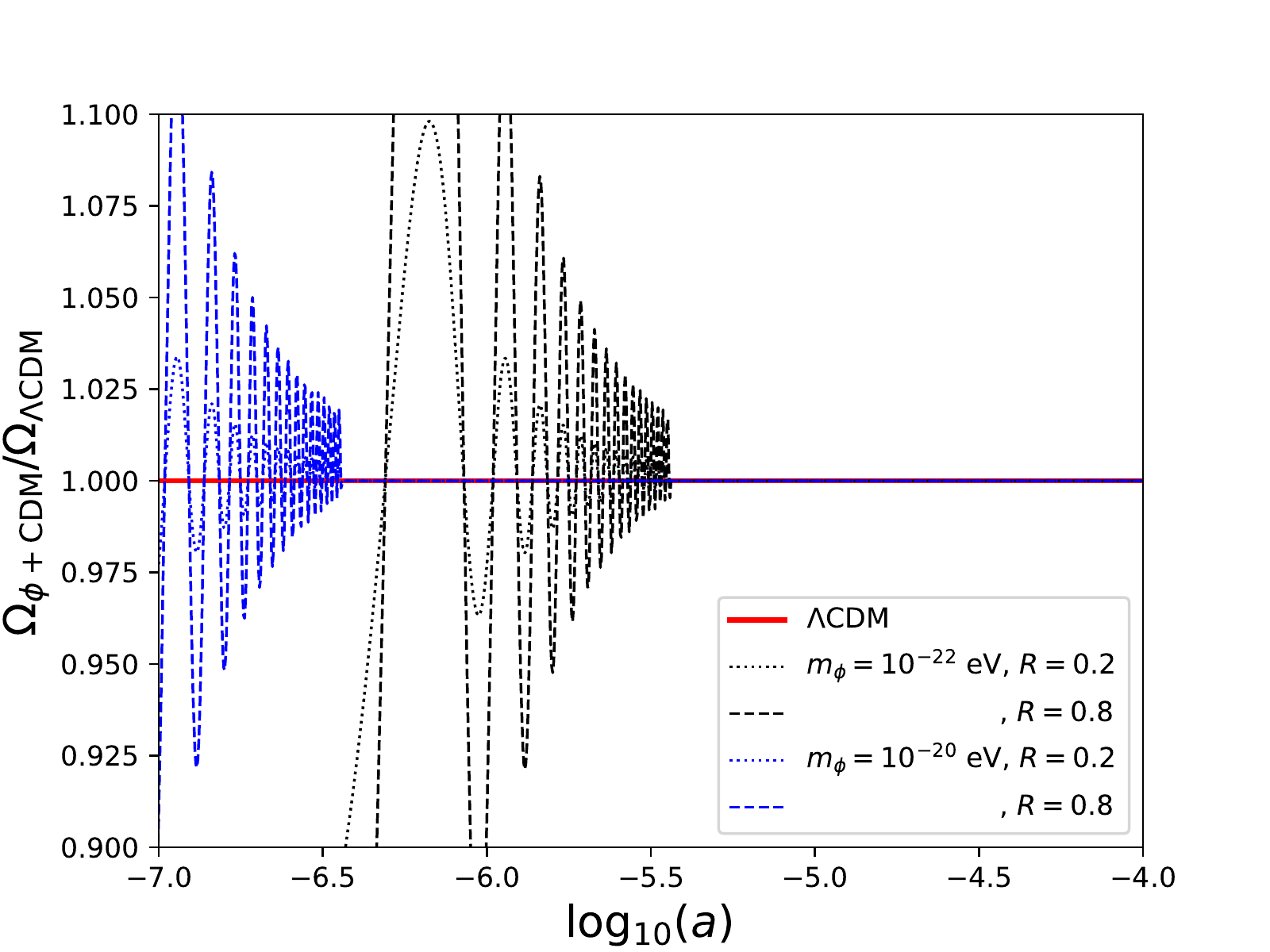}
    \includegraphics[trim = 0mm  0mm 0mm 0mm, clip, width=10.5cm, height=7.5cm]{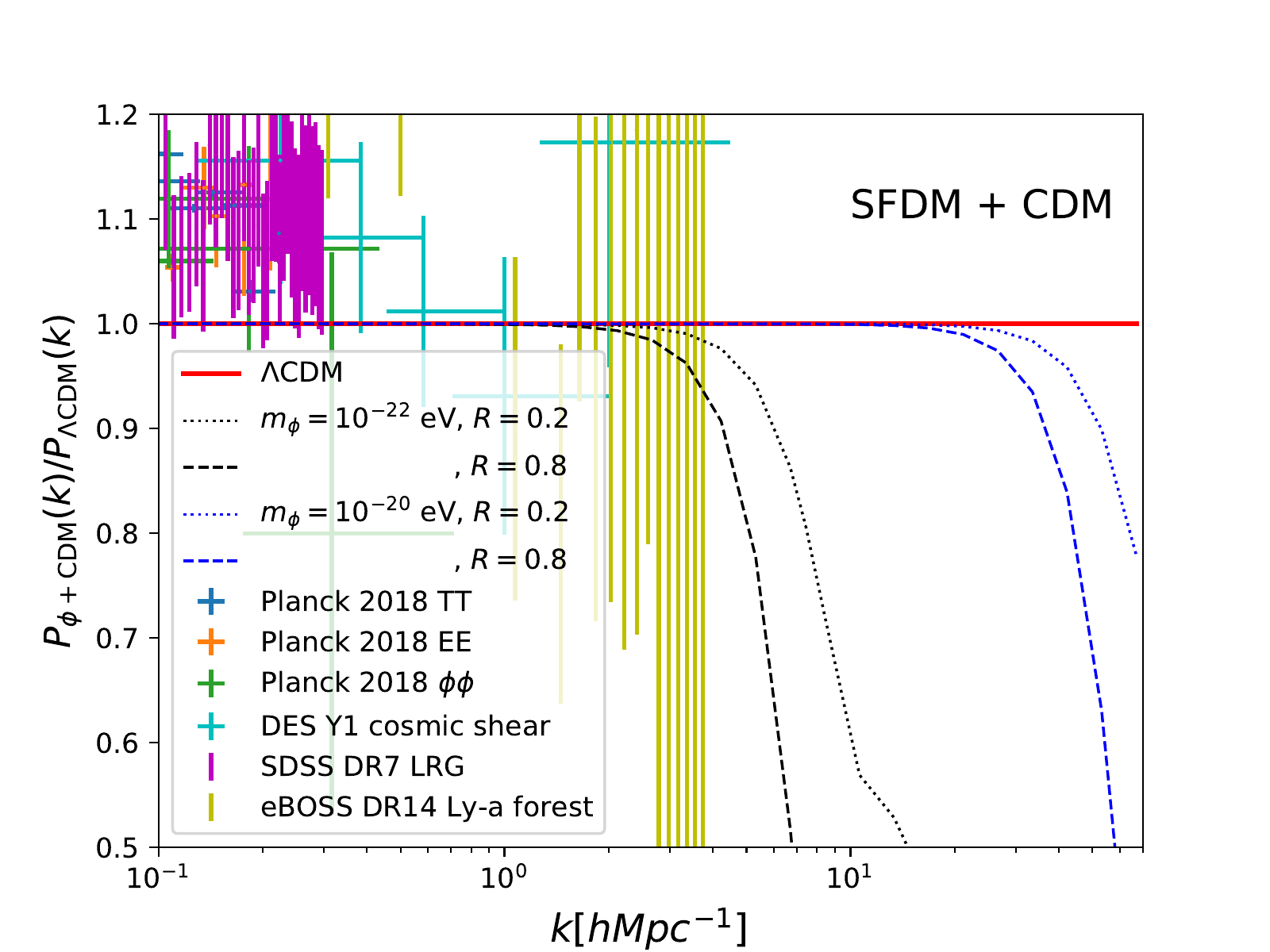}
    }
    \caption{\footnotesize{Evolution, \lo{zoomed-in on early times}, of the ratio of the density parameters $\Omega_{\phi   + \mathrm{CDM}}/\Omega_{\Lambda\mathrm{CDM}}$ (left) and the ratio of the linear matter power spectrum (right) at $z=0$, for a  SFDM + CDM model using $\Lambda\mathrm{CDM}$ as reference (solid red lines). The field potential is the quadratic one $V(\phi) = \frac{1}{2}m_{\phi}^2\phi^2$, and $R$ represents the ratio of the field contribution to the total DM. Black lines represent a mass value of $m_{\phi} = 10^{-22}$ eV while blue lines refer to $m_{\phi} = 10^{-20}$ eV. Dashed and dotted lines represent $R = 0.8$ and $R = 0.2$ values respectively. The data points for the MPS were obtained from~\cite{Chabanier:2019eai}.}}
    \label{fig:numerical_SFDM_CDM}
\end{figure*}

In the first combination we assume the field one has $V(\phi_1) = \frac{1}{2}m_{\phi 1}^2\phi_1^2$ while the rest of the dark matter density is conformed by CDM, therefore $\Omega_{\phi 2,0}=0$. 
As we can see on the left panel of Figure \ref{fig:numerical_SFDM_CDM}, the main difference in the background 
is the start of the scalar field oscillations due to the mass and $R$ values. If $R>0.5$ the oscillations approach those of a single scalar field while if $R<0.5$ the oscillations are less evident until they disappear when $R=0$ (there is only CDM). 
In the mass power spectrum, the right panel of the same figure, we notice 
the expected cut-off in small scales for the lightest mass values ($m_{\phi 1}<10^{-20}$ eV)  but the behaviour is lost for heavier masses ($m_{\phi 1}>10^{-20}$ eV) because the scalar field behaves like dust over the shown scales. 
Similar to the background, we found that the MPS of the combinations is bounded by the cases $R=1$ (only SFDM) and $R=0$ (only CDM). 
We can see an example of this in Figure \ref{fig:numerical_SFDM_CDM} for $m_{\phi 1} = 10^{-22}$ eV (black line) and $10^{-20}$ eV (blue line) with $R=0.2$ (dotted) and $R=0.8$ (dashed). For $R=0.8$ the oscillations approach the case of a single SFDM 
but disappear for $R=0.2$, this is evident for $m_{\phi 1} = 10^{-22}$ eV. In the same way, in the right panel, when $R=0.8$ the mass power spectrum shows a cut-off similar to the single case for both values $m_{\phi 1} = 10^{-22}$ and $10^{-20}$ eV, but it behaves like CDM (red solid line) for $R=0.2$. 
\\

\begin{figure*}
    \centering
    \makebox[14cm][c]{
    \includegraphics[trim = 0mm  0mm 10mm 0mm, clip, width=10.cm, height=7.5cm]{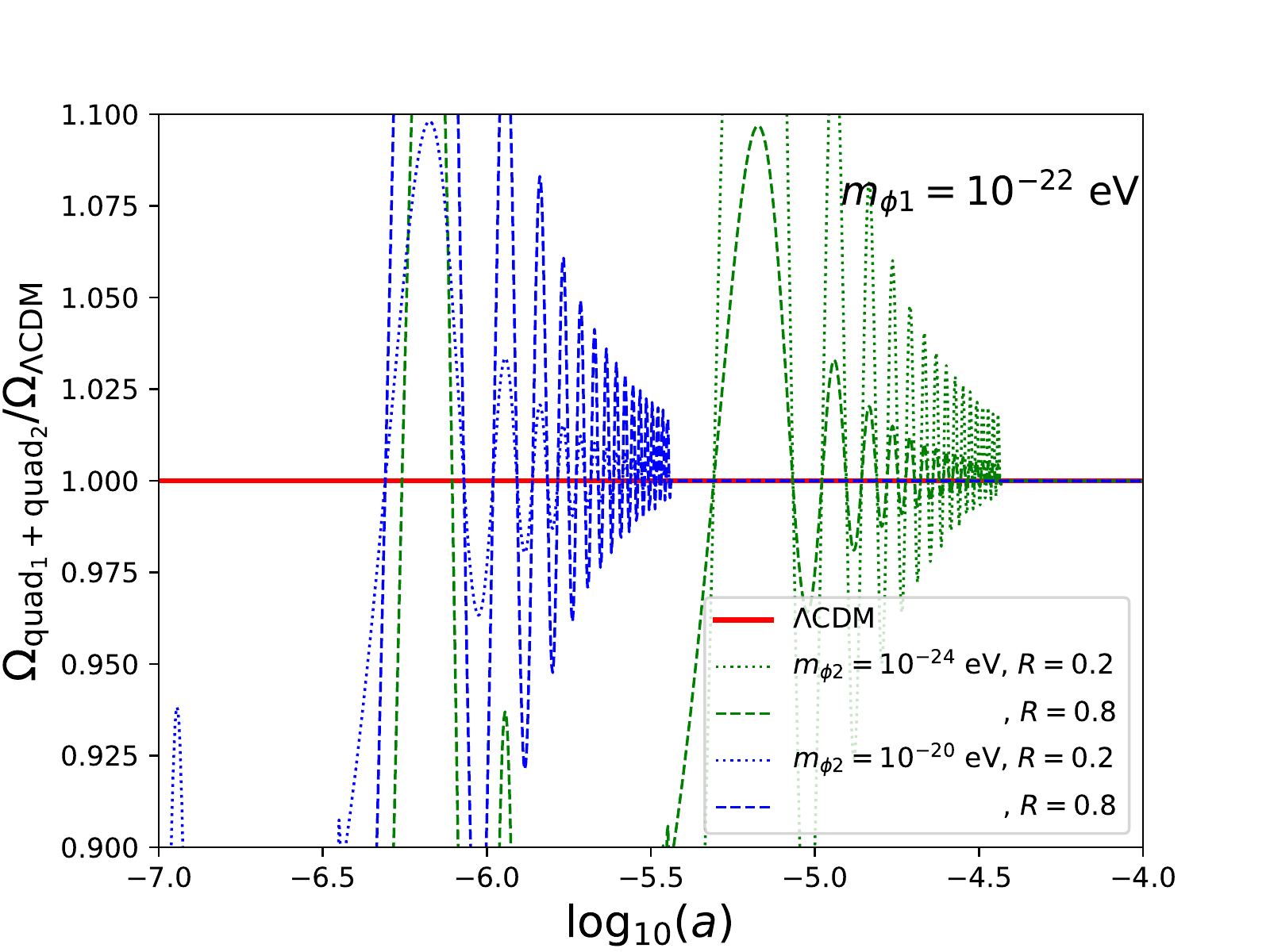} 
    \includegraphics[trim = 0mm  0mm 0mm 0mm, clip, width=10.5cm, height=7.5cm]{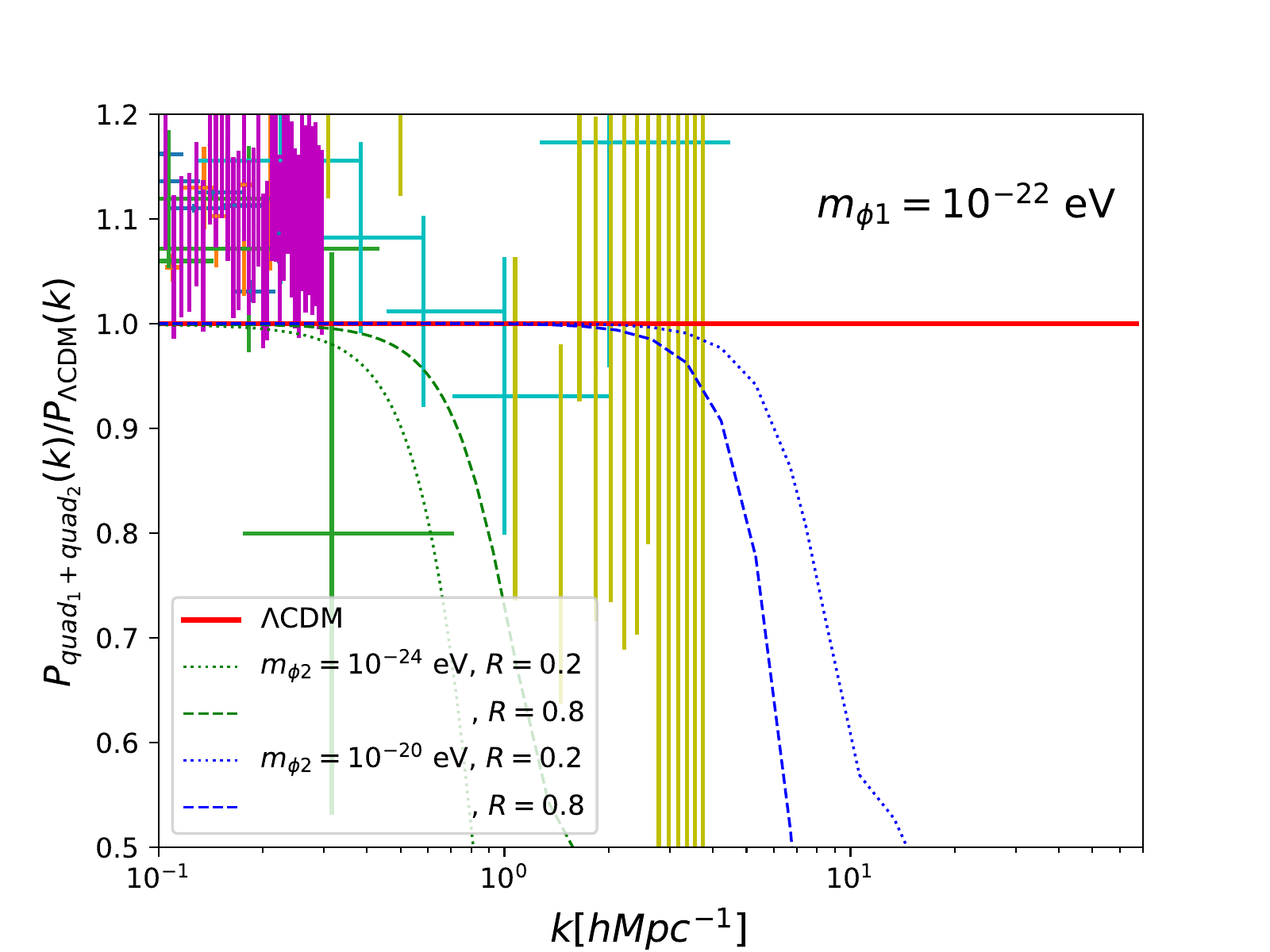}
    }
    \caption{\footnotesize{Evolution, \lo{zoomed-in on early times}, of the ratio of the density parameters $\Omega_{quad_1 + quad_2}/\Omega_{\Lambda\mathrm{CDM}}$ (left) and the ratio of the linear matter power spectrum (right) at $z=0$, for a double field model using $\Lambda\mathrm{CDM}$ as reference (solid red lines). The potential for both fields is the quadratic one $V(\phi) = \frac{1}{2}m_{\phi}^2\phi^2$, and $R$ represents the ratio of the fields contribution to the total DM. Green lines represent a mass value of $m_{\phi 2} = 10^{-24}$ eV and blue lines refer to $m_{\phi 2} = 10^{-20}$ eV while $m_{\phi 1}$ is fixed to $10^{-22}$ eV. Dashed and dotted lines represent $R = 0.8$ and $R = 0.2$ values respectively. The data points for the MPS were obtained from~\cite{Chabanier:2019eai}. MPS data labels are the same as in Figure \ref{fig:numerical_SFDM_CDM}.}}
    \label{fig:numerical_quad_quad}
\end{figure*}

For the second combination we assume two scalar fields with no CDM, 
both of them with a quadratic potential $V(\phi_{1,2}) = \frac{1}{2}m_{\phi 1,2}^2\phi_{1,2}^2$ 
(see Figure \ref{fig:numerical_quad_quad}). We found that the total contribution to the background evolution, in particular the oscillations of the fields, depends on the contribution of each one through $R$ and are bounded by the oscillations of the lightest and the heaviest field respectively. For the MPS, 
the cut-off is more evident when the lightest field accounts for the principal contribution to $\Omega_{\rm DM,0}$ 
($m_{\phi 2} > m_{\phi 1}$ and $R>0.5$ or $m_{\phi 2}<m_{\phi 1}$ and $R<0.5$), the behavior is  closer to single field models ($R=0$ or $R=1$) depending of the dominant field, as expected. 
In the left panel of Figure \ref{fig:numerical_quad_quad} we show the total contribution of the fields to the background using $m_{\phi 1} = 10^{-22}$ eV with $m_{\phi 2} = 10^{-24}$ eV (green). The oscillations are more evident for $R=0.8$ (dashed) than for $R=0.2$ (dotted) and both of them are bounded by the single cases (solid). This is seen as well, but it is less noticeable due to the masses values, for the combination $m_{\phi 1} = 10^{-22}$ eV with $m_{\phi 2} = 10^{-20}$ eV (blue). For the MPS, we use the same values and we find a behavior similar to the previous case. 
\\

\begin{figure*}
    \centering
    \makebox[14cm][c]{
    \includegraphics[trim = 0mm  0mm 10mm 0mm, clip, width=10.cm, height=7.5cm]{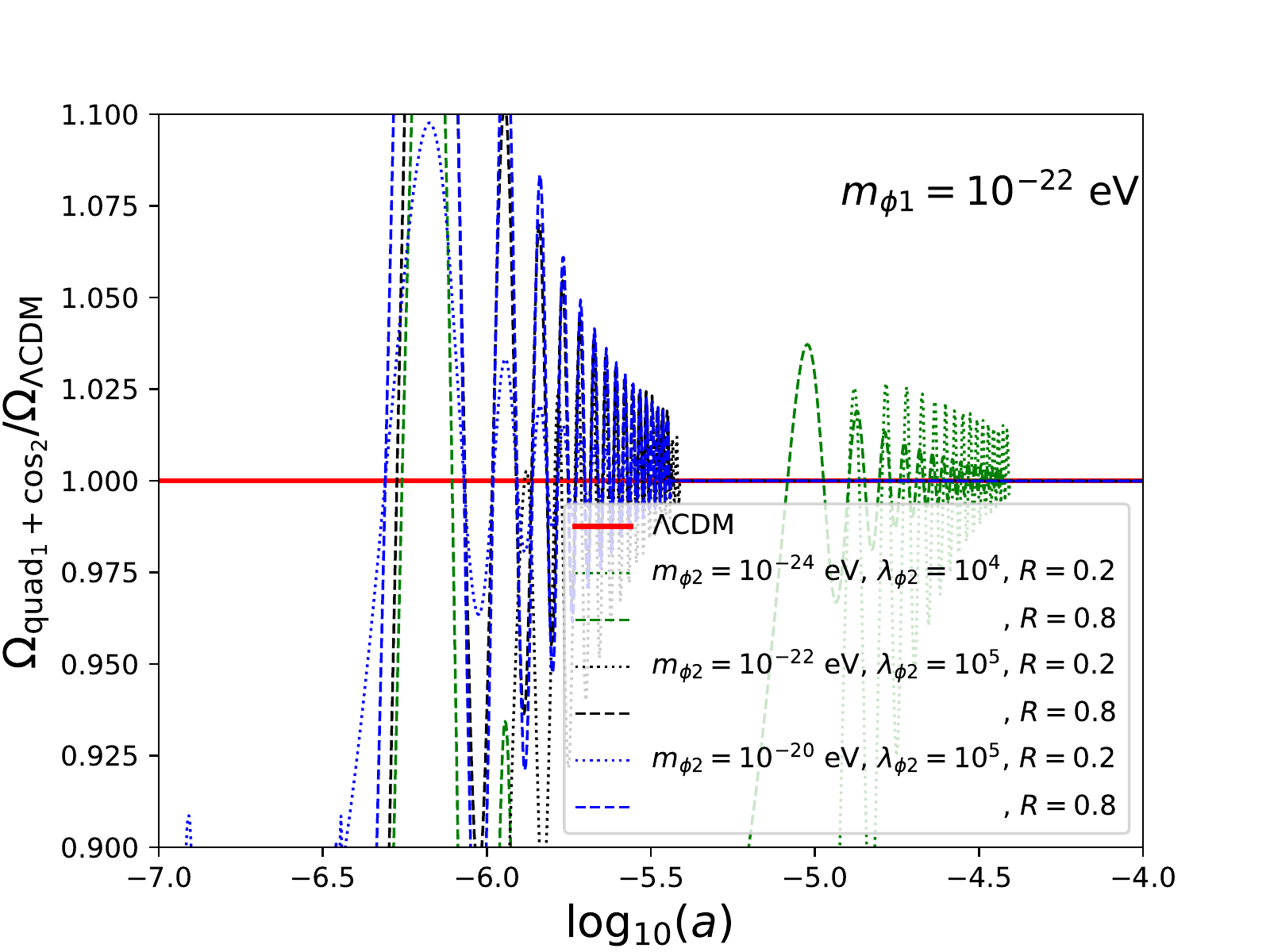}
    \includegraphics[trim = 0mm  0mm 0mm 0mm, clip, width=10.5cm, height=7.5cm]{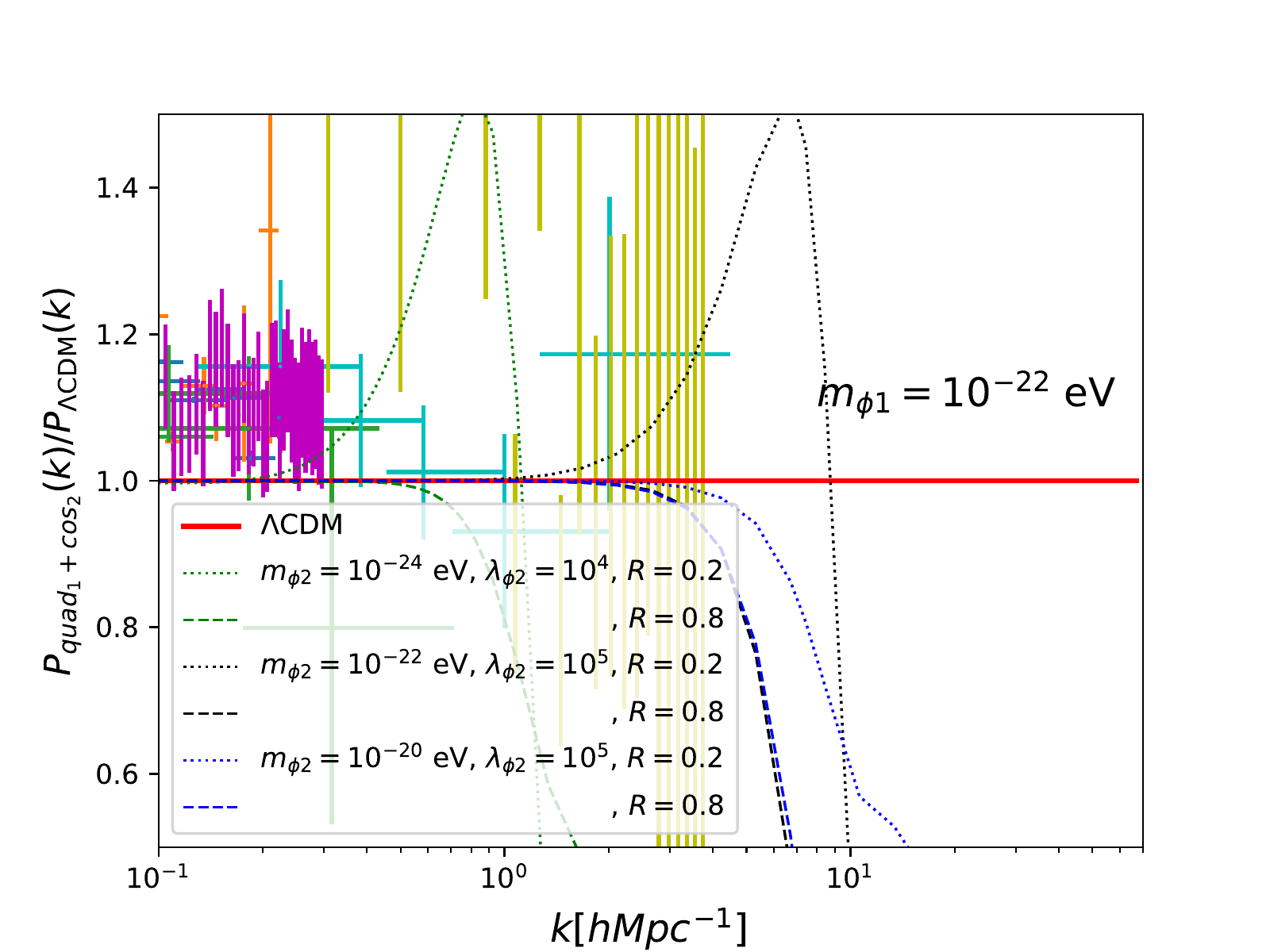}
    }
    \caption{\footnotesize{Evolution, \lo{zoomed-in on early times}, of the ratio of the density parameters $\Omega_{quad_1 + cos_2}/\Omega_{\Lambda\mathrm{CDM}}$ (left) and the ratio of the linear matter power spectrum (right) at $z=0$, for a model with two fields using $\Lambda\mathrm{CDM}$ as reference (solid red lines). Here, the field one has the potential $V(\phi) = \frac{1}{2}m_{\phi}^2\phi^2$ with mass value $m_{\phi} = 10^{-22}$ eV as reference. While the second field has $V(\phi_2) = m_{\phi 2}^2f^2\left[1+\cos(\phi_2/f)\right]$. Green lines represent a mass value of $m_{\phi 2} = 10^{-24}$ eV with $\lambda_{\phi 2} = 10^4$, black lines indicate $m_{\phi 2} = 10^{-22}$ eV with $\lambda_{\phi 2} = 10^5$ and blue lines refer to $m_{\phi 2} = 10^{-20}$ eV with $\lambda_{\phi 2} = 10^5$. Dashed and dotted lines represent $R = 0.8$ and $R = 0.2$ values respectively, where $R$ represents the ratio of the fields contribution to the total DM. The data points for the MPS were obtained from~\cite{Chabanier:2019eai}. MPS data labels are the same as in Figure \ref{fig:numerical_SFDM_CDM}.}}
    \label{fig:numerical_quad_cos}
\end{figure*}

We also study a third  combination of double field but now with $V(\phi_1) = \frac{1}{2}m_{\phi 1}^2\phi_1^2$ and $V(\phi_2) = m_{\phi 2}^2f^2\left[1+\cos(\phi_2/f)\right]$ respectively. For the background we observe a similar behavior than in the previous combinations, 
that is, once we kept the masses fixed and solely modify the $\lambda_{\phi 2}$ values, we see no difference among them.
The oscillations of the fields depend on the masses and $R$, similarly to the quadratic potentials case. See, for example, Figure \ref{fig:numerical_quad_cos} where we take $m_{\phi 1} = 10^{-22}$ eV along with  $m_{\phi 2} = 10^{-24}$ eV with $\lambda_{\phi 2} = 10^4$ (green), $m_{\phi 2} = 10^{-22}$ eV with $\lambda_{\phi 2} = 10^5$ (black) and $m_{\phi 2} = 10^{-20}$ eV with $\lambda_{\phi 2} = 10^5$ (blue) for $R=0.2$ (dotted) and $R=0.8$ (dashed). For the MPS we see  differences depending on $m_{\phi 1,2}$ values or $R$ and also by varying $\lambda_{\phi 2}$ values. In order to see a bump in the MPS  we need that the second field dominates ($R<0.5$) with lighter mass and higher $\lambda_{\phi 2}$. In the right side of Figure \ref{fig:numerical_quad_cos} we can see the bump before the cut-off ($k\sim 0.9$) 
for the green and black dotted lines, because in both cases the second field dominates and $\lambda_{\phi 2}$ has high values with lighter masses. On the other hand, in the blue dotted line we do not observe the bump yet because the field becomes too  heavy ($m_{\phi 2}\geq10^{-20}$ eV). 
\\

\begin{figure*}
    \centering
    \makebox[14cm][c]{
    \includegraphics[trim = 0mm  0mm 10mm 0mm, clip, width=10.cm]{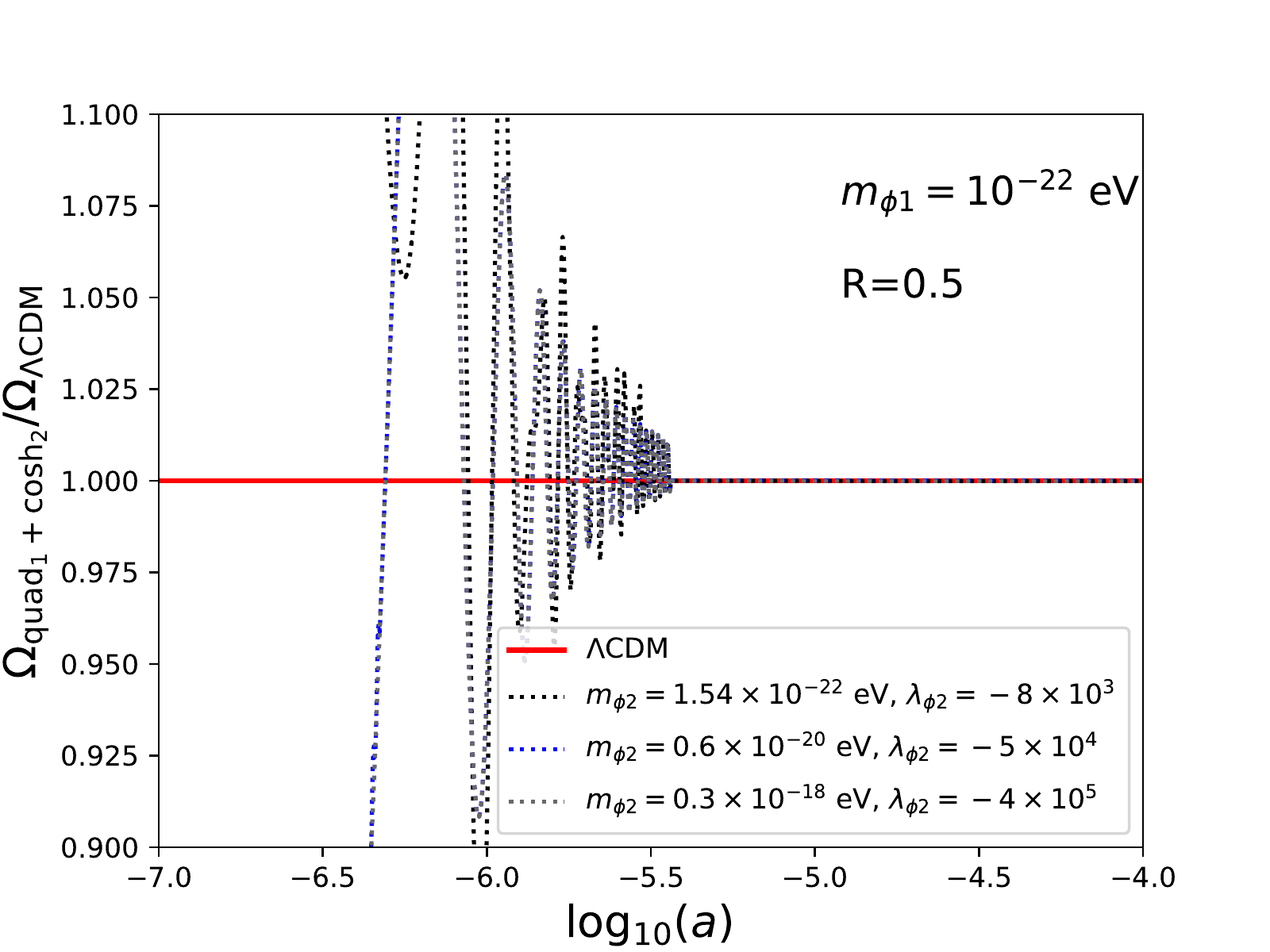}
    \includegraphics[trim = 0mm  0mm 0mm 0mm, clip, width=10.5cm]{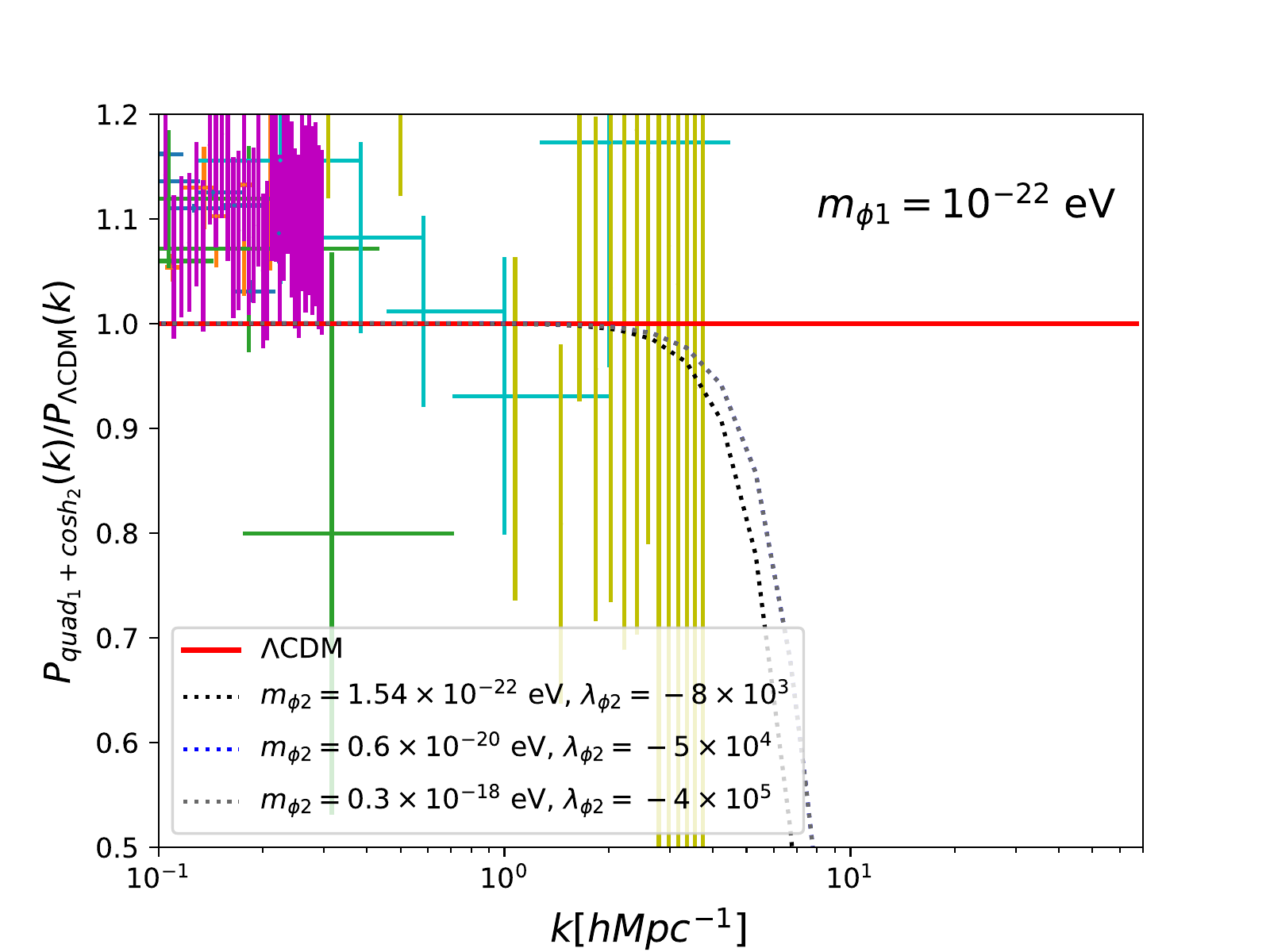}
    }
    \caption{\footnotesize{Evolution, \lo{zoomed-in on early times}, of the ratio of the density parameters $\Omega_{quad_1 + cosh_2}/\Omega_{\Lambda\mathrm{CDM}}$ (left) and the ratio of the linear matter power spectrum (right) at $z=0$, for a model with two fields using $\Lambda\mathrm{CDM}$ as reference (solid red lines). The field one has the potential $V(\phi) = \frac{1}{2}m_{\phi}^2\phi^2$ with mass value $m_{\phi} = 10^{-22}$ eV as reference, and the second field has $V(\phi_2) = m_{\phi 2}^2f^2\left[\cosh(\phi_2/f)-1\right]$. Black lines represent a mass value of $m_{\phi 2} = 1.54 \times 10^{-22}$ eV with $\lambda_{\phi 2} = -8\times10^3$, blue lines indicate $m_{\phi 2} = 0.6\times10^{-20}$ eV with $\lambda_{\phi 2} = -5\times10^4$ and green lines refer to $m_{\phi 2} = 0.3\times10^{-18}$ eV with $\lambda_{\phi 2} = -4\times10^5$. Dotted lines represent $R = 0.5$ where $R$ represents the ratio of the fields contribution to the total DM. The data points for the MPS were obtained from~\cite{Chabanier:2019eai}. MPS data labels are the same as in Figure \ref{fig:numerical_SFDM_CDM}.}}
    \label{fig:numerical_quad_cosh}
\end{figure*}

Finally, we study the combination of the field potential $V(\phi_1) = \frac{1}{2}m_{\phi 1}^2\phi_1^2$ with $V(\phi_2) = m_{\phi 2}^2f^2\left[\cosh(\phi_2/f)-1\right]$. In the background and MPS we have a similar behavior as before. This can be seen in Figure \ref{fig:numerical_quad_cosh} where we take the combination $m_{\phi 1} = 10^{-22}$ eV, $m_{\phi 2} = 1.54 \times 10^{-22}$ eV, $\lambda_{\phi 2} = -8\times10^3$ (black), $m_{\phi 1} = 10^{-22}$ eV, $m_{\phi 2} = 0.6 \times 10^{-20}$ eV, $\lambda_{\phi 2} = -5\times10^4$ (blue) and $m_{\phi 1} = 10^{-22}$ eV, $m_{\phi 2} = 0.3 \times 10^{-18}$ eV, $\lambda_{\phi 2} = -4\times10^4$ (gray). 
It is important to note that both fields have the same contribution ($R = 0.5$), other combinations require further analysis. 
In the mass power spectrum, at small scales, we see that $\lambda_{\phi 2}$ has a greater contribution 
compared to the previous case because in the cosh-like potential the scalar field mass depends on the $\lambda_{\phi 2}$ value, contrary to the cos-like potential in which these parameters are independent. When $m_{\phi 2} \sim m_{\phi 1}$ we can appreciate a different behavior depending on the value of $\lambda_{\phi}$, for example, the black dotted line. However, if $m_{\phi 2}\geq m_{\phi 1}$, the change in $\lambda_{\phi 2}$ does not display noticeable changes (blue dotted and gray dotted lines). 

In general for the background, we found \lo{slight differences at early times with respect to $\Lambda$CDM, where the oscillations presented could give us information about how light the scalar fields masses can be.} We found too a cut-off at small scales in the mass power spectrum that differentiates our model from the CDM, and the shape below the cut-off depends on the multi-field dynamics.  


\begin{figure*}[t!]
	\centering
	 \makebox[\textwidth][c]{
	\includegraphics[trim = 0mm  0mm 10mm 0mm, clip, width=11cm]{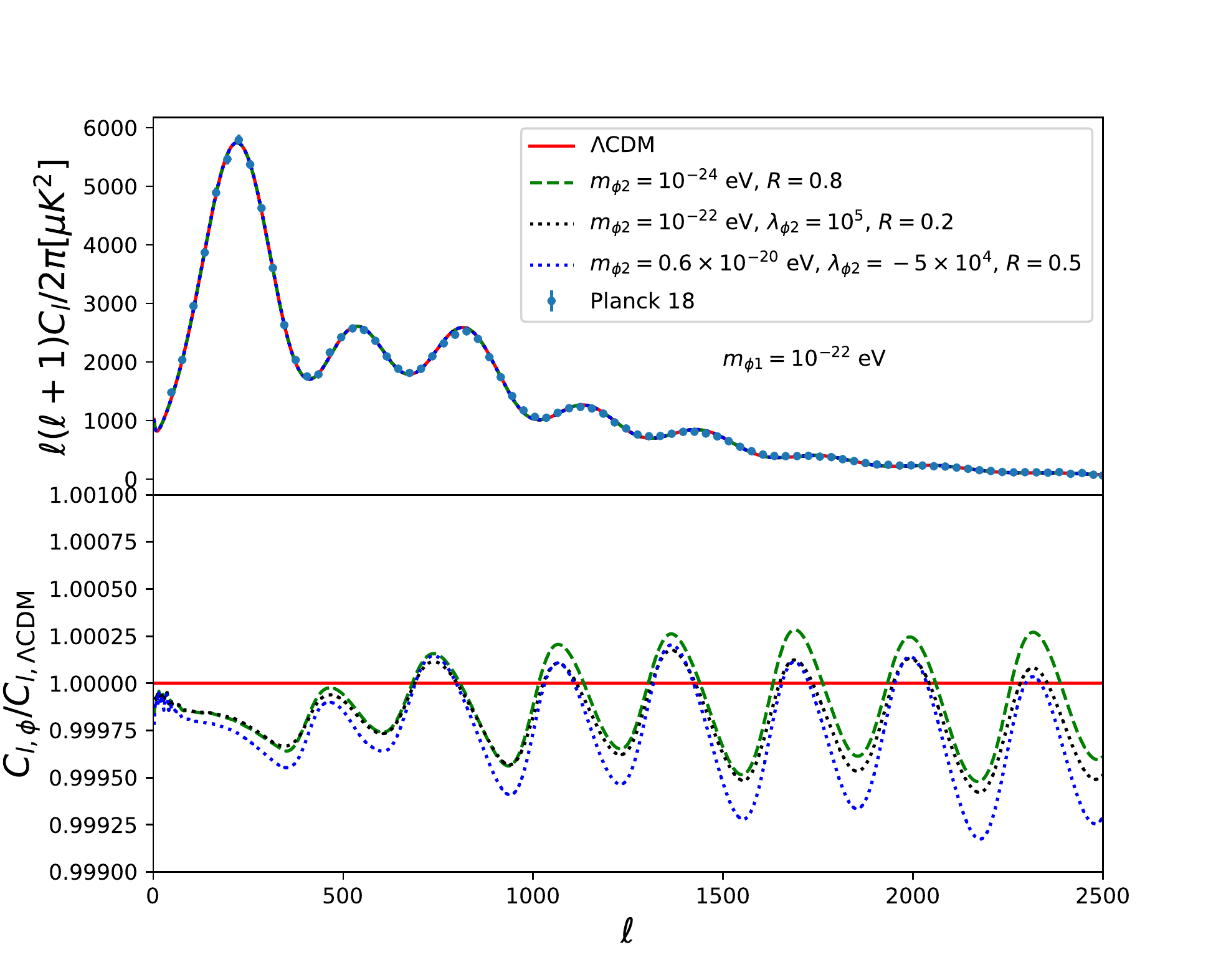}}
	\caption{\footnotesize{The CMB power spectrum (upper panel) and the ratio (lower panel) using $\Lambda\mathrm{CDM}$ as reference (solid red lines), for $V(\phi_{1,2}) = \frac{1}{2}m_{\phi 1,2}^2\phi_{1,2}^2$ (green dashed line), $V(\phi_1) = \frac{1}{2}m_{\phi 1}^2\phi_1^2$ with $V(\phi_2) = m_{\phi 2}^2f^2\left[1+\cos(\phi_2/f)\right]$ (black dotted line) and $V(\phi_1) = \frac{1}{2}m_{\phi 1}^2\phi_1^2$ with $V(\phi_2) = m_{\phi 2}^2f^2\left[\cosh(\phi_2/f)-1\right]$ (blue dotted line).}} 
	\label{fig:CMB}
\end{figure*}

Regarding the CMB spectrum, when the mass of at least one of the fields, for any combination, is less than $m_{\phi i}<10^{-26}$ eV, the spectrum of the fields differs from the $\Lambda$CDM spectrum as seen in \cite{Hlozek:2014lca}. However, for masses greater than $m_{\phi i}>10^{-26}$ eV the same CMB power spectrum is obtained as for $\Lambda$CDM regardless of the value of $\lambda_{\phi}$ or $R$. See, for example Figure \ref{fig:CMB} where we show the CMB power spectra for $V(\phi_{1,2}) = \frac{1}{2}m_{\phi 1,2}^2\phi_{1,2}^2$ with $m_{\phi 1}=10^{-22}$ eV, $m_{\phi 2} = 10^{-24}$ eV and $R=0.8$ (green dashed line); $V(\phi_1) = \frac{1}{2}m_{\phi 1}^2\phi_1^2$ with $V(\phi_2) = m_{\phi 2}^2f^2\left[1+\cos(\phi_2/f)\right]$ for $m_{\phi 1}=10^{-22}$ eV, $m_{\phi 2} = 10^{-22}$ eV, $\lambda_{\phi 2} = 10^5$ and $R=0.2$ (black dotted line) and $V(\phi_1) = \frac{1}{2}m_{\phi 1}^2\phi_1^2$ with $V(\phi_2) = m_{\phi 2}^2f^2\left[\cosh(\phi_2/f)-1\right]$ for $m_{\phi 1}=10^{-22}$ eV, $m_{\phi 2} \propto 10^{-20}$ eV, $\lambda_{\phi 2} = -5\times10^4$ and $R=0.5$ (blue dotted line). For the three cases we have obtained nearly the same CMB power spectra as for the $\Lambda$CDM model (solid red line). We also plotted the Planck-18 dataset as reference \cite{Planckdata}.

\section{Comparison with data}
\label{Constraints}
In this section we present the Bayesian inference procedure in order to constrain the MSFDM models by using different data sets. In order to perform the cosmological analysis, we generate Monte Carlo Markov Chains with the Metropolis-Hastings algorithm using a modified version of the Monte Python code \cite{Brinckmann:2018cvx, montepython} with the modified version of CLASS used in subsection \ref{numerical}. We verified that our chains converged using the Gelman-Rubin criterion $R-1 < 0.03$ implemented in Monte Python. Then we discuss the merits of the models with respect to CDM within the framework of the Bayesian model selection. See \cite{Padilla:2019mgi} for a bayesian inference review.

\subsection{Cosmological constraints}
\label{cosmological}
In the previous section we saw that the main difference throughout the models rests on the mass power spectrum at small scales, hence we use the 3D matter power spectrum inferred from Lyman-$\alpha$ data from BOSS and eBOSS collaboration \cite{Chabanier:2019eai}. We also use the Ly-$\alpha$ BAO from eBOSS DR14 \cite{Cuceu:2019for}, the Galaxy BAO from DR12 \cite{Alam:2016hwk}, 6dFGS \cite{6dFGS} and SDSS DR7 \cite{Ross:2014qpa}, and the SNe Ia  survey Pantheon \cite{Scolnic:2017caz} to improve the constraining power. As mentioned in \cite{Chabanier:2019eai}, the process of inferring the MPS is model-dependent, therefore we will consider the constraints obtained here as an approximation. On the other hand, we found that the parameter constraints describing the scalar-field remain the same when incorporating the Planck 18 dataset. For this reason and to reduce computational time we have only used the data already described. See appendix \ref{ap:triangle} for further details.

The sampling parameters in the analysis are the physical baryon density parameter $\omega_{b,0}$, the logarithmic power spectrum scalar amplitude $\log\left(10^{10}A_s\right)$, the scalar spectral index $n_s$, the Thomson scattering optical depth due to reionization $\tau_{reio}$, the scalar field masses $m_{\phi i}$, the decay parameters $\lambda_{\phi i}$ and, instead of cold dark matter density parameter, we use the scalar field density parameters $\Omega_{\phi i, 0}$. However we found, from previous analysis, that the posteriors of $\log\left(10^{10}A_s\right)$, $n_s$ and $\tau_{reio}$ do not present a change respect to $\Lambda$CDM, therefore we will kept these parameters fixed. See appendix \ref{ap:triangle}. The flat priors used for the remaining sampling parameters are as follow: $H_0 = [10,100]$ for the Hubble constant in $\mathrm{km}$ $\mathrm{s^{-1} Mpc^{-1}}$, $\omega_{b,0} = [0.005,0.1]$ for the physical baryon density and $\Omega_{\phi 1, 0} = \Omega_{\phi 2, 0} = [0,1]$ for the scalar field density parameters today. For the scalar field parameters, we choose the priors to be consistent with the numerical results we found with CLASS, and since these parameters can take values of powers of ten, we chose a logarithmic base to efficiently cover the entire parameter space and reduce the computational cost. This means that we have $\log_{10}\left(m_{\phi i}/ \mathrm{eV}\right) = [-24,-17]$ for the scalar field masses, $\log_{10}\left(\lambda_{\phi i}\right) = [1,6]$ for the parameter on the trigonometric cosine potential and $\lambda_{\phi i, \mathrm{aux}} = [-6,-1]$ for the hyperbolic cosine potential. For this case, we cannot use directly $\log_{10}\left(\lambda_{\phi i}\right)$ because $\lambda_{\phi i}<0$, so in order to cover the parameter space we use the auxiliary variable $\lambda_{\phi i, \mathrm{aux}}$ whose relation with $\lambda_{\phi i}$ is given by $\lambda_{\phi i} = -10^{-\lambda_{\phi i, \mathrm{aux}}}$. In what follows we will use $\mathrm{quad}_i$ to refer to the potential $V_i(\phi_i) = (1/2) m^2_{\phi i} \phi^2_i$, $\mathrm{cos}_i$ for $V_i(\phi_i) = m^2_{\phi i} f^2_i \left[1 + \cos(\phi_i/f_i) \right]$ and $\mathrm{cosh}_i$ for $V_i(\phi_i) = m^2_{\phi i} f^2_i \left[\cosh(\phi_i/f_i) -1 \right]$. If there is no subscript, it means that it is the single field case. 
\\

\begin{figure*}[t!]
	\centering
	\makebox[11cm][c]{
	\includegraphics[width=4.5cm, height=5.cm]{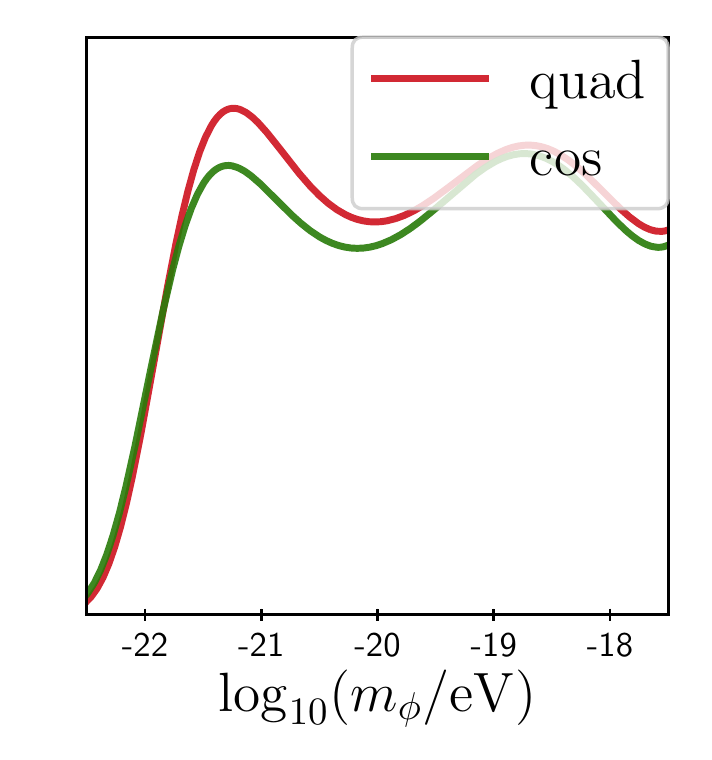}
	\includegraphics[width=4.5cm, height=5.cm]{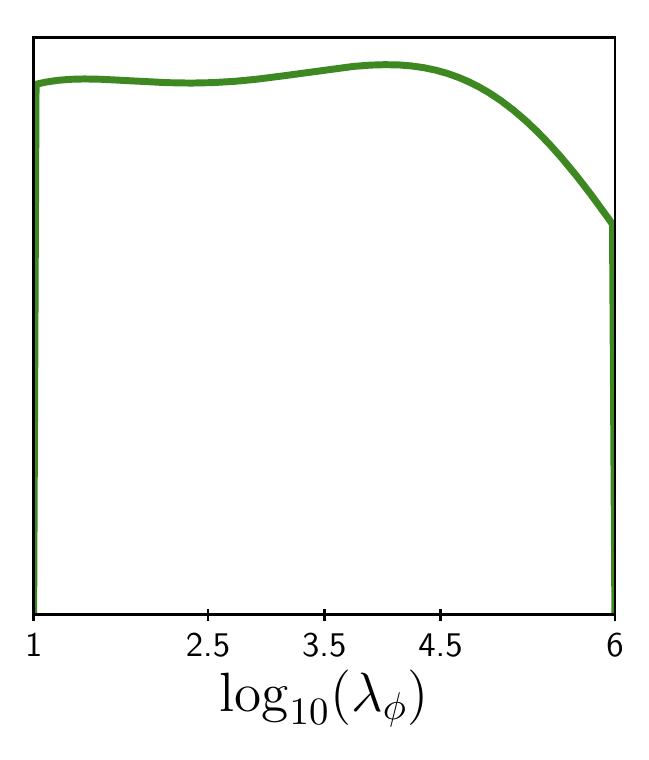}
	\includegraphics[width=4.5cm, height=5.cm]{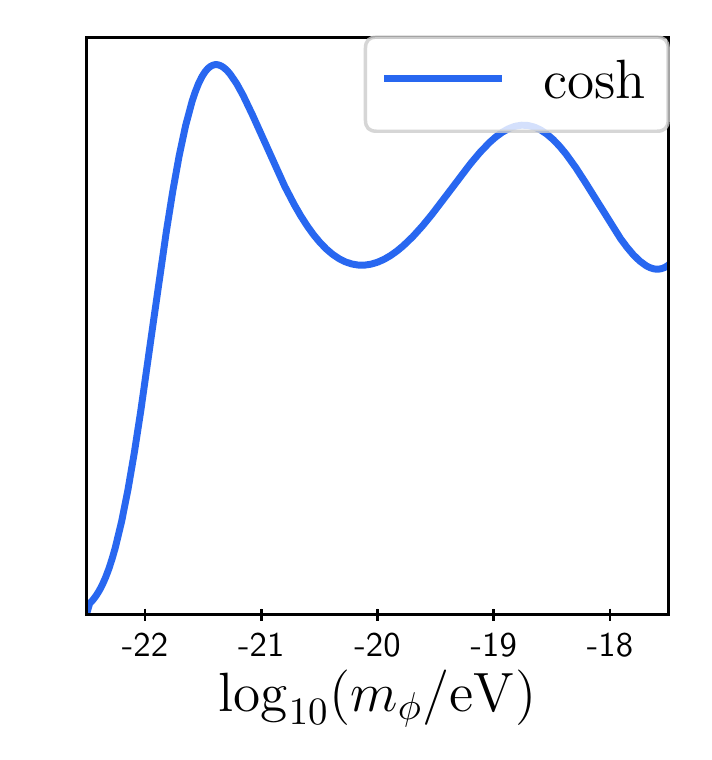}
	\includegraphics[width=4.5cm, height=5.cm]{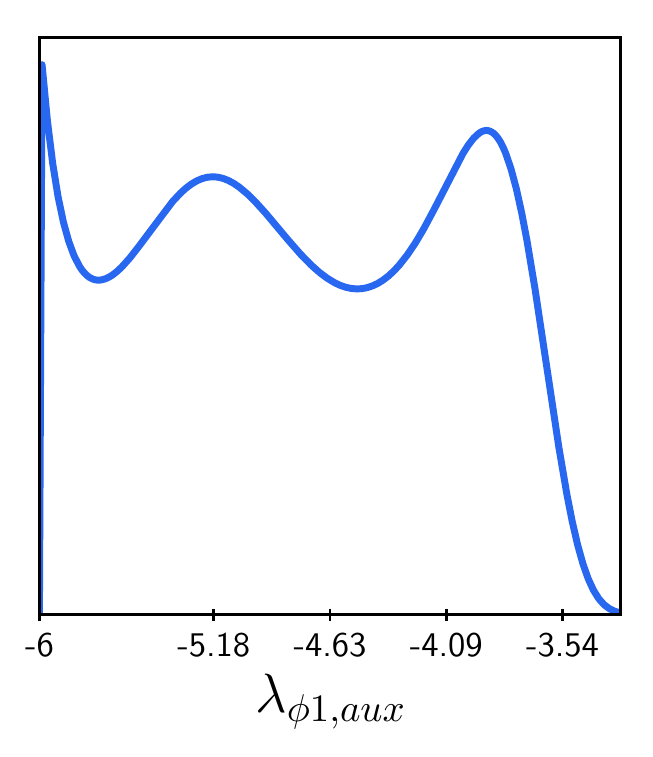}
	}
	\caption{\footnotesize{1D posterior distribution for the single SFDM with potentials $V(\phi) = 1/2 m_{\phi}^2 \phi^2$ (red), $V(\phi) = m_{\phi}^2f^2\left[1+\cos(\phi/f)\right]$ (green) and $V(\phi) = m_{\phi}^2f^2\left[\cosh(\phi/f)-1\right]$ (blue). The constraints on the field mass are very similar for the three potentials, whereas the self-interaction parameter $\lambda_\phi$ of the trigonometric potential remains unconstrained by the data. }}
	\label{SFDM_1D}
\end{figure*}

\begin{figure*}[t!]
	\centering
	\includegraphics[width=0.47\textwidth]{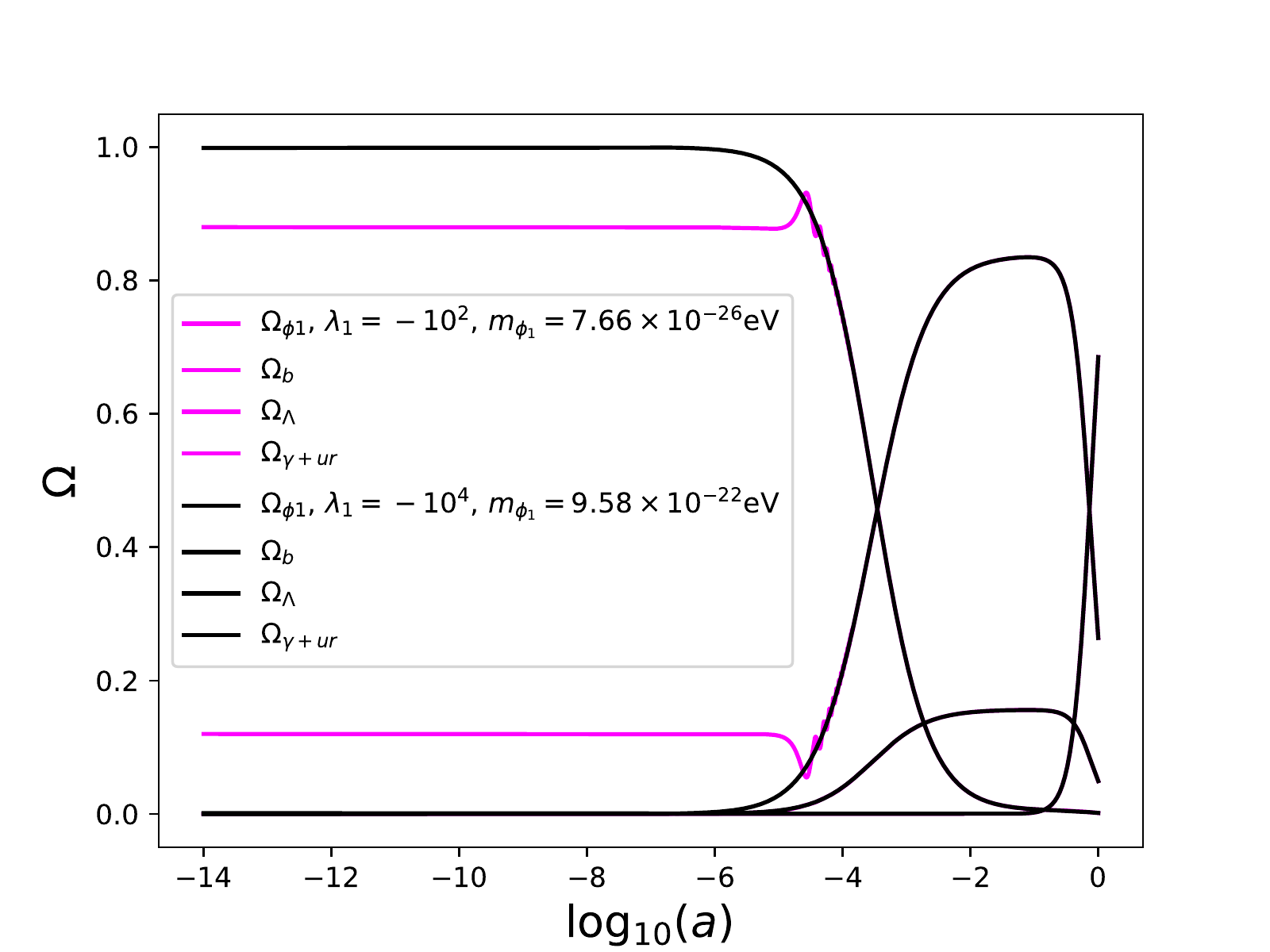}
	\includegraphics[width=0.47\textwidth]{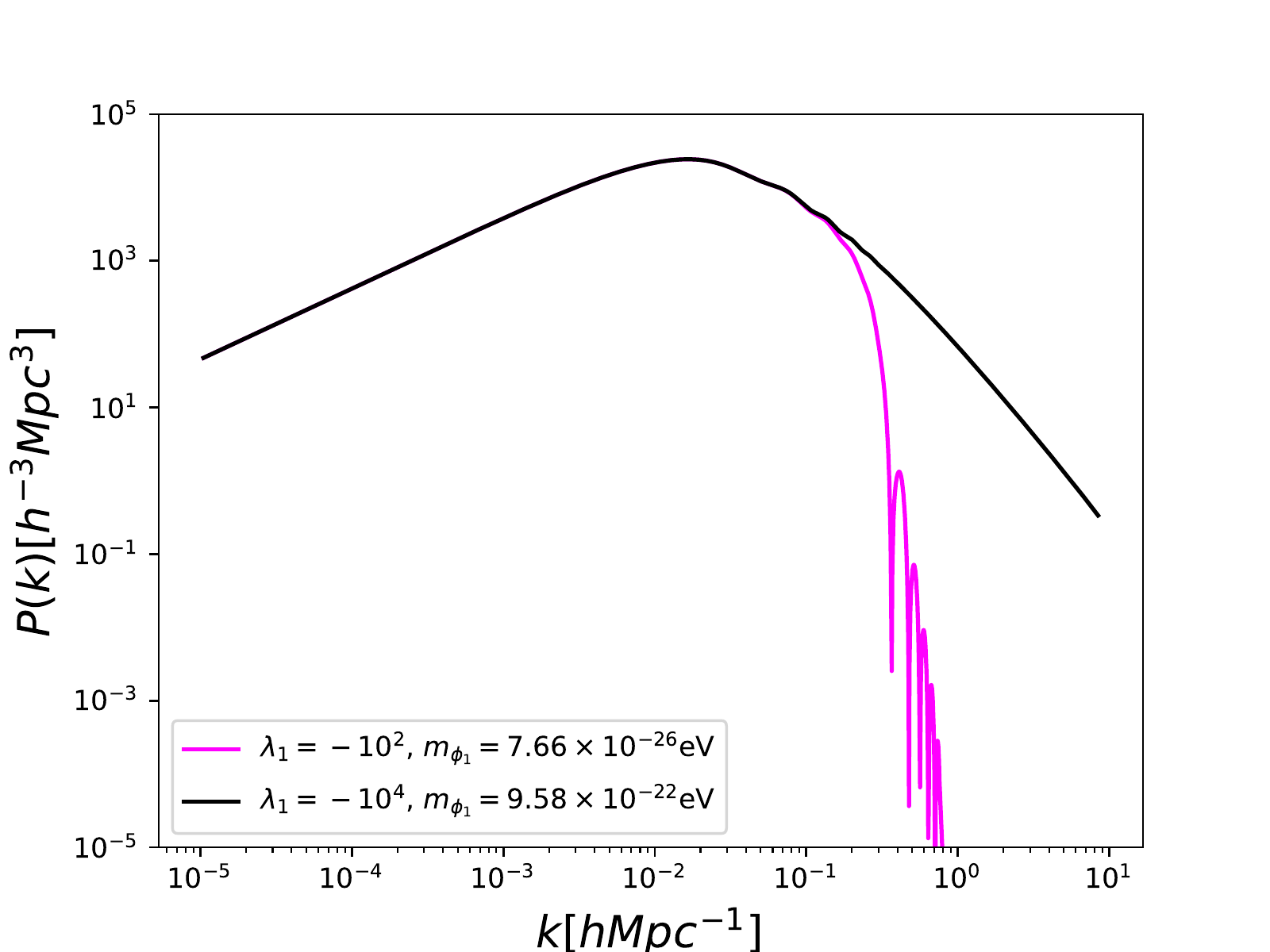}
	\caption{\footnotesize{Background and MPS evolution for the SFDM with potential $V(\phi) = m_{\phi}^2f^2\left[\cosh(\phi/f)-1\right]$. Magenta lines correspond to $\lambda_{\phi} = -10^2$ with $m_{\phi} = 7.66 \times 10^{-26}$ eV, which are ruled out from the constraints obtained. We can see the scalar field contribution to radiation. On the other hand, black lines, that correspond to $\lambda_{\phi} = -10^4$ with $m_{\phi} = 9.58 \times 10^{-22}$ eV, represent values within the confidence regions. }}
	\label{SFDM_cosh_Omegas}
\end{figure*}

First, in Figure~\ref{SFDM_1D}, we show the constraints for the single field cases, that is, the 1D marginalized posterior distribution  of the free parameters for each model corresponding to the potentials $V(\phi) = 1/2 m_{\phi}^2 \phi^2$, $V(\phi) = m_{\phi}^2f^2\left[1+\cos(\phi/f)\right]$, and $V(\phi) = m_{\phi}^2f^2\left[\cosh(\phi/f)-1\right]$. It can be noticed that the posteriors of the field masses are very similar for the three potentials, and that the observations considered can only put a lower bound on it, which is $\log (m_\phi/\mathrm{eV}) \gtrsim -21.9$ at $95\%$ CL for the quadratic and hyperbolic cosine potentials and $\log (m_\phi/\mathrm{eV}) \gtrsim -21.8$ at $95\%$ CL for the trigonometric cosine potential. In the case of the trigonometric potential, we see that its extra parameter $\lambda_\phi$ appears unconstrained and the posterior looks practically the same with the prior we considered above for this parameter, which is consistent with previous studies~\citep{LinaresCedeno:2020dte,Armengaud:2017nkf,Irsic:2017yje}. As for the hyperbolic cosine, we have already mentioned the close relationship between the field mass $m_\phi$ and the self-interaction parameter $\lambda_\phi$, and then the constraints on the former are translated to $\log(-\lambda_{\phi}) \gtrsim 3.6$ at $95\%$ CL. The excluded values of $\lambda_\phi$ correspond to the cases in which SFDM has a significant contribution as an early radiation component. An example of this can be seen in Figure \ref{SFDM_cosh_Omegas}.

\begin{figure*}[t!]
	\centering
	\makebox[11cm][c]{
	\includegraphics[width=5.cm, height=4.cm]{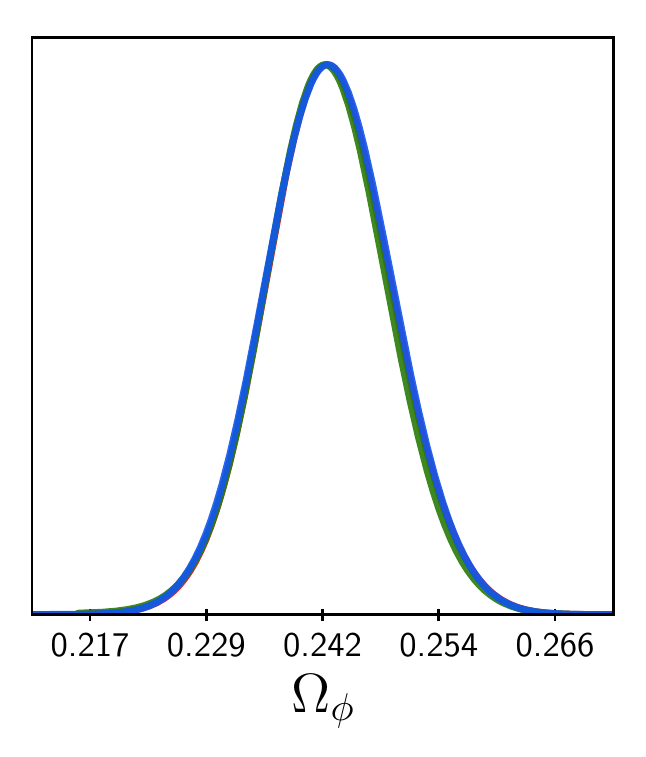}
	\includegraphics[width=5.cm, height=4.cm]{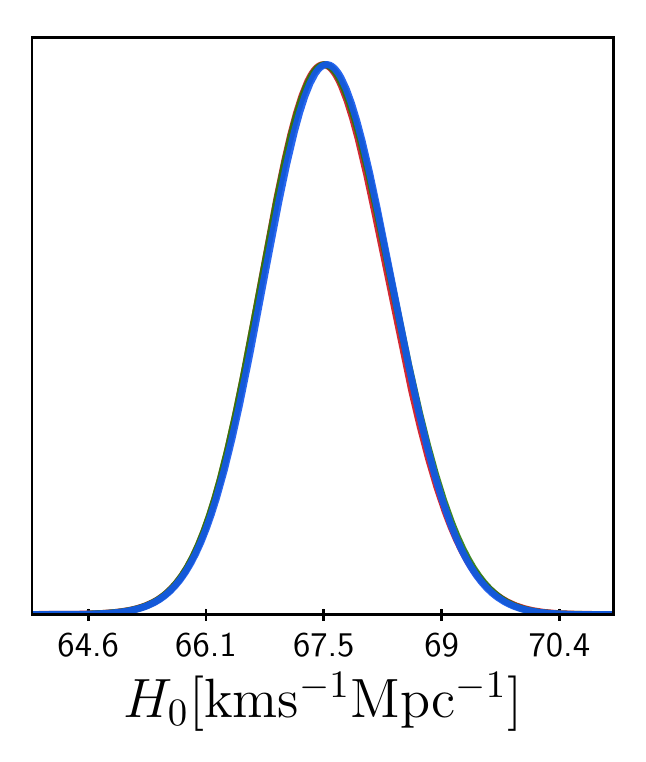}
	\includegraphics[width=5.cm, height=4.cm]{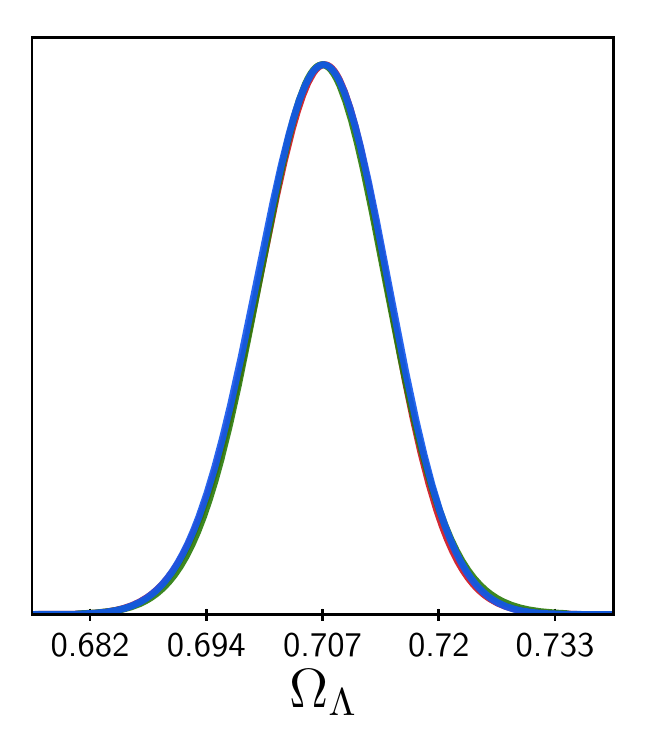}
	}
	\caption{\footnotesize{1D posterior distribution for the cosmological common parameters of all the models studied in this work.}}
	\label{fig:common_params}
\end{figure*}

For completeness, we also present in Figure \ref{fig:common_params} the 1D posterior distributions for the cosmological common parameters of the models: the density parameters (SFDM  and $\Lambda$), and the Hubble constant $H_0$. The posteriors are very similar for the three models, which show that one recovers the results of the standard $\Lambda$CDM model.

\begin{figure*}[t!]
	\centering
	\makebox[11cm][c]{
	\includegraphics[width=5.cm, height=4.2cm]{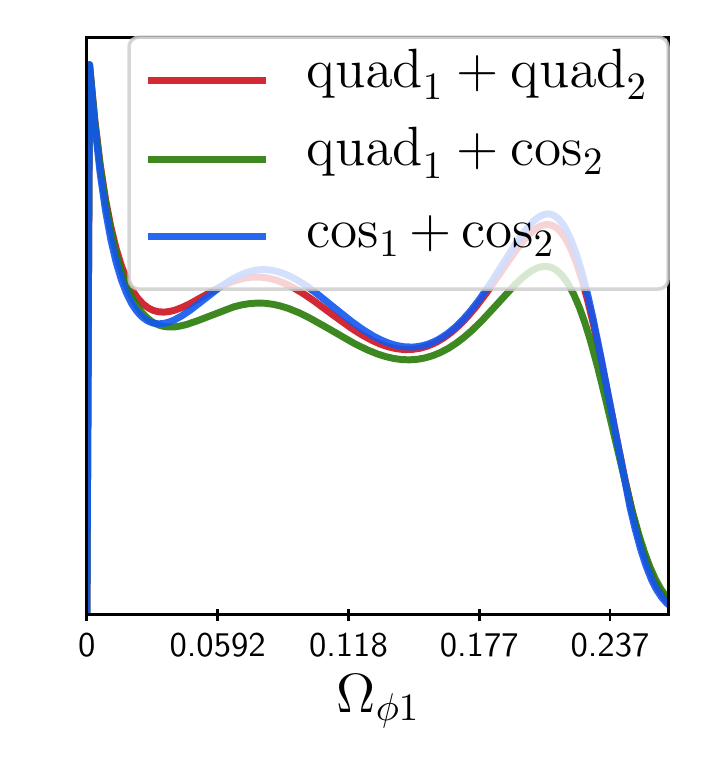}
	\includegraphics[width=5.cm, height=4.2cm]{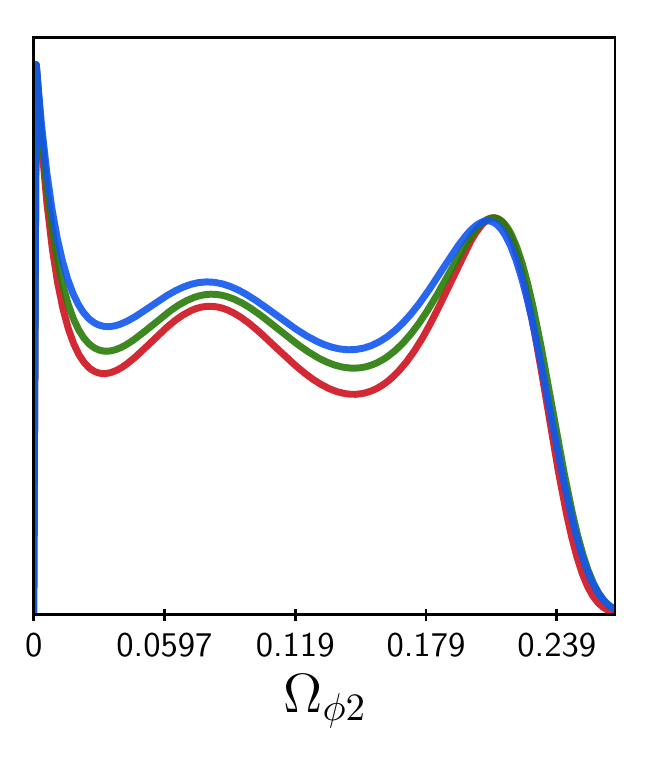}
	\includegraphics[width=5.cm, height=4.2cm]{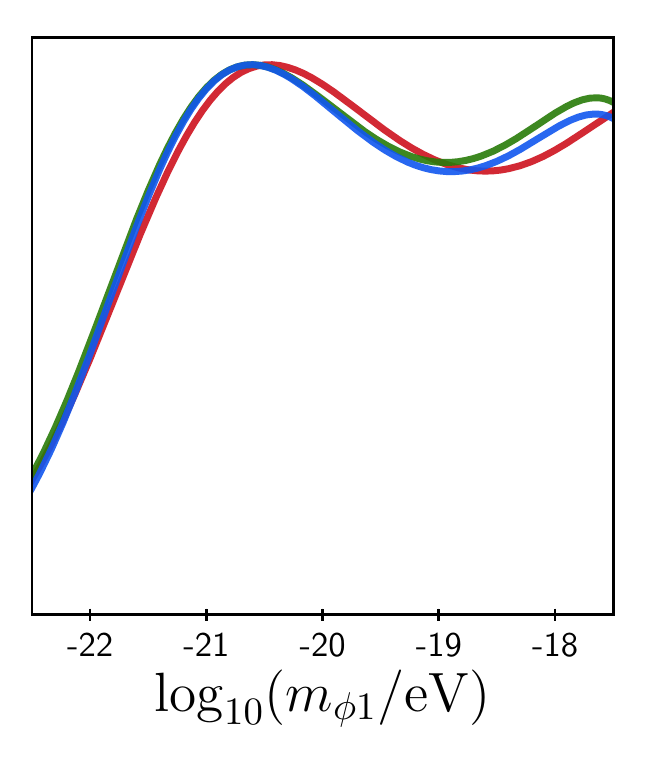}
	}
	\makebox[11cm][c]{
	\includegraphics[width=5.cm, height=4.2cm]{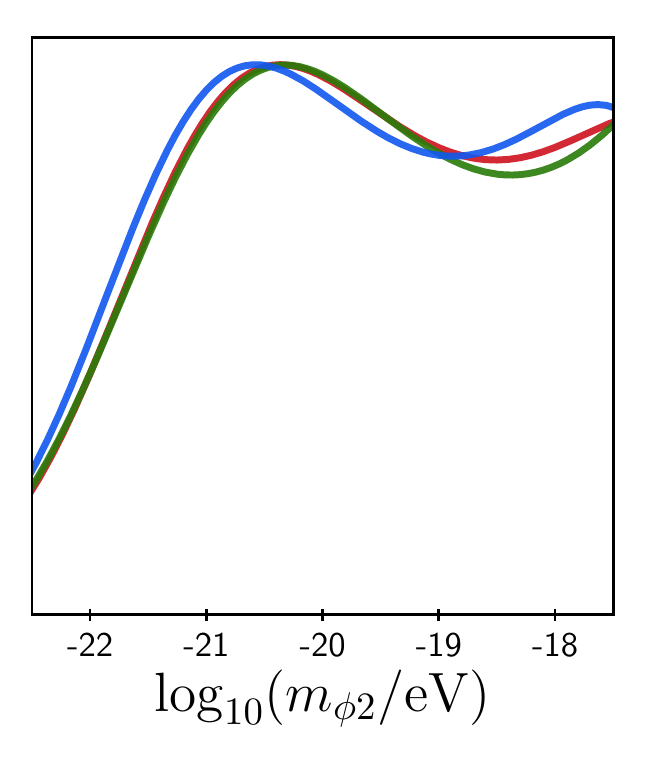}
	\includegraphics[width=5.cm, height=4.2cm]{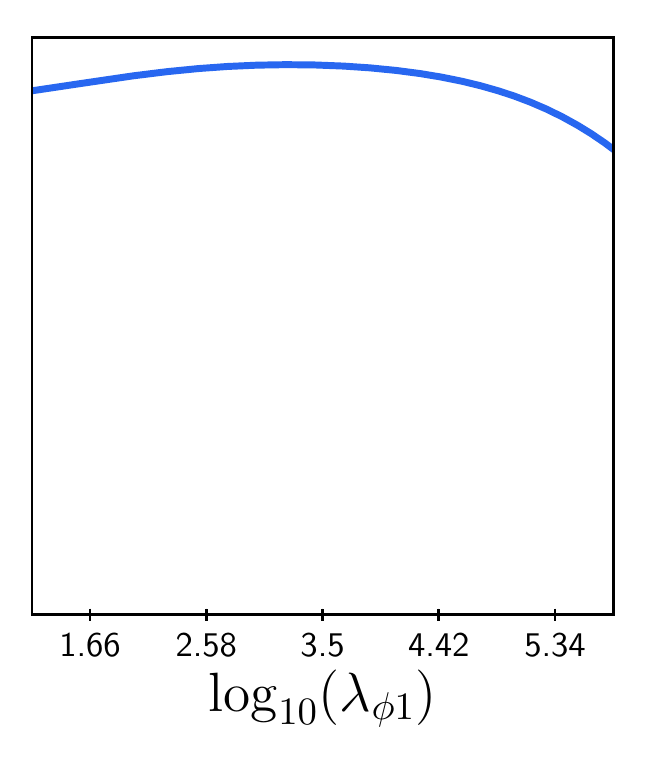}
	\includegraphics[width=5.cm, height=4.2cm]{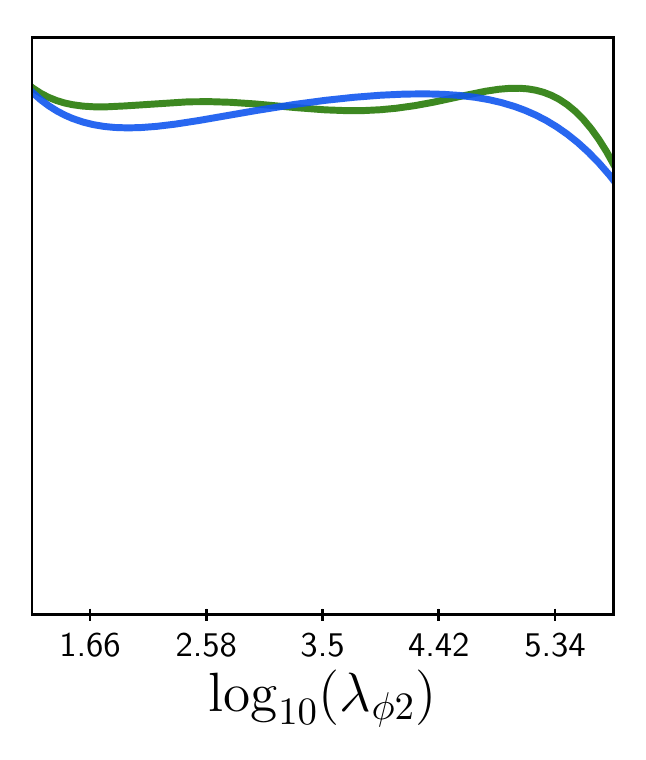}
	}
	\caption{\footnotesize{1D posterior distribution for the two SFDM with the potentials $V(\phi_{1,2}) = 1/2 m_{\phi 1,2}^2 \phi_{1,2}^2$ (red), $V(\phi_1) = 1/2 m_{\phi 1}^2 \phi_1^2$ with $V(\phi_1) = m_{\phi 1}^2f^2\left[1+\cos(\phi_1/f)\right]$ (green) and $V(\phi_{1,2}) = m_{\phi 1,2}^2f^2\left[1+\cos(\phi_{1,2}/f)\right]$ (blue).}}
	\label{twoSFDM_1D}
\end{figure*}

In Figure \ref{twoSFDM_1D} we show the posteriors for the combinations $V(\phi_1) = 1/2 m_{\phi 1}^2 \phi_1^2$ with $V(\phi_2) = 1/2 m_{\phi 2}^2 \phi_2^2$, $V(\phi_1) = 1/2 m_{\phi 1}^2 \phi_1^2$ with $V(\phi_2) = m_{\phi 2}^2f^2\left[1+\cos(\phi_2/f)\right]$ and $V(\phi_{1,2}) = m_{\phi 1,2}^2f^2\left[1+\cos(\phi_{1,2}/f)\right]$.
For the quadratic-quadratic and quadratic-cos combinations we found a lower bound for the mass of both fields given by $\log_{10} (m_{\phi 1,2}/\mathrm{eV}) = -22.1$ at $95\%$. While, for the cos-cos combination, the lower bound for the mass of field one is $\log_{10} (m_{\phi 1}/\mathrm{eV}) = -22.1$ at $95\%$ and for the second field $\log_{10} (m_{\phi 2}/\mathrm{eV}) = -22.2$ at $95\%$.  We also found that the density parameter $\Omega_{\phi i}$ of one field depends on the $\Omega_{\phi j}$ of the other field. That is, if $\Omega_{\phi 1}$ dominates then $\Omega_{\phi 2}$ is small, and on the other way around because the total contribution remains constant. In \cite{Broadhurst:2018fei} the authors propose two scalar fields (with the possibility of a third one) with masses $m_{\phi 1} \approx 10^{-22}$ eV and $m_{\phi 2} \approx 10^{-20}$ eV, and these values are within our confidence regions. The fact that the lighter field dominates over the massive field is also in agreement with our results. The triangle plots of the MSFDM parameters in Figure \ref{SFDM_1D} and Figure \ref{twoSFDM_1D} can be seen in appendix \ref{ap:triangle}. 

As we have mentioned, the main objective is to present the model of Multiple Scalar Fields as Dark Matter. Although, for completeness, we compute the Bayes factor, $\ln \mathcal{B}_{i, j}$, by using the numerical package MCEvidence \cite{2017arXiv170403472H}, and the ratio is done respect to CDM. The results are shown in Table~\ref{BayesFactor}. Following the Jeffreys guideline \cite{Padilla:2019mgi}, if $\ln \mathcal{B}_{i, j} > 5$ we have a decisive strength against model $i$; if $5 >\ln \mathcal{B}_{i, j} > 2.5$ it means a strong strength; if $2.5 > \ln \mathcal{B}_{i, j} > 1$ we have a significant strength and if $\ln \mathcal{B}_{i, j} < 1$ the data prefers model $i$.
In general, we should be careful taking the values of the Bayes factor since some of the parameters were unconstrained by the data \cite{Trotta:2008qt}.
On the other hand, the fact that the posteriors of the parameters of the scalar fields are not Gaussian prevents us from using other information criteria such as the AIC or the BIC; or else, the Bayesian complexity, since they need (or assume) the Gaussianity of the posterior. See, for example, \cite{Liddle:2007fy} or \cite{Kunz:2006mc}. On the other hand, we do not present the Bayes factor for the trigonometric potential cosine nor for the models with more than one field because the posteriors are not mono-modal and some parameters are not constrained which can cause the method used in MCEvidence to fail \cite{2017arXiv170403472H}.

We found that the maximum of the likelihoods correspond to the same value $-\ln \mathcal{L}_{\rm max} = 523.03$. That is, we can find a suitable combination of parameters such that the model in place resembles the behaviour of a cold dark matter.

%
\begin{table*}[ht!]
\begin{center}
\begin{tabular}{|p{5em}|p{4em}|p{4em}|p{7em}|p{7em}|p{7em}|p{6em}|}
\hline
& $\mathrm{quad}$ & $\cosh$ & $\mathrm{quad}_1+\mathrm{quad}_2$  \\
\hline
$\ln \mathcal{B}_{i, \Lambda{\rm CDM}}$ & 1.59 & 0.85 & 2.13   \\
\hline
\end{tabular}
\caption{\footnotesize{Bayes factor for SFDM and double-SFDM with different potentials using $\Lambda$CDM as reference.}}
\label{BayesFactor}
\end{center}
\end{table*}

\section{Conclusions}

In this paper we presented the cosmological constraints of the Multi Scalar Field Dark Matter (MSFDM) model, in which we assume the dark matter is composed of different ultralight scalar fields. The idea was introduced to alleviate some of  the cosmological and astrophysical  discrepancies, for instance, the distinct values of the single scalar field mass obtained when considering different observations, where more than one field is needed in order to explain them. As a first approximation, we took for granted that the scalar fields are real, spatially homogeneous and they do not interact with each other. We presented the equations that describe the evolution of the background and perturbations for $N$ scalar fields, and by using 
the polar change of variables we avoided the scalar field characteristic oscillations. Thus, we obtained a general expression for the fields evolution that depend on the potential and its derivatives, in particular, the equations for the potentials  $V(\phi) = 1/2 m_{\phi}^2 \phi^2$, $V(\phi) = m_{\phi}^2f^2\left[1+\cos(\phi/f)\right]$ and $V(\phi) = m_{\phi}^2f^2\left[\cosh(\phi/f)-1\right]$. Under this change of variables, these three configurations are described by a single system of equations where the parameter $\lambda_{\phi}$ is able to decide the type of potential to use.

We showed the evolution for the background, mass power spectrum (MPS) and the CMB power spectrum using a modified version of the CLASS code \cite{class, Urena-Lopez:2015gur, Cedeno:2017sou} with two dark matter components. We considered the following combinations: a) cold dark matter and a scalar field with potential $V(\phi) = 1/2 m_{\phi}^2 \phi^2$, and two scalar field models, with b) both potentials $V(\phi) = 1/2 m_{\phi}^2 \phi^2$, c)  $V(\phi) = 1/2 m_{\phi}^2 \phi^2$ and $V(\phi) = m_{\phi}^2f^2\left[1+\cos(\phi/f)\right]$, and d) $V(\phi) = 1/2 m_{\phi}^2 \phi^2$ and $V(\phi) = m_{\phi}^2f^2\left[\cosh(\phi/f)-1\right]$. Since the fields are independent from each other, we introduced the parameter $R$ to have the field one as reference.
We showed that the evolution of the background is mainly affected at the beginning of the scalar fields oscillations since its amplitude depends on the masses and the contribution that each one has to the total dark matter, being bounded by the heaviest and lightest masses; however these do not depend on $\lambda_{\phi}$. After the oscillations have started, the fields evolution only depend on $\Omega_{\phi i,0}$. On the other hand we found that, in all combinations, the MPS presents the characteristic cut-off of the single field at small scales. However, it does not present the oscillations of the single case due to the superposition of the different fields. This cut-off is far form the $\Lambda$CDM behavior when the lightest field dominates, and if the heaviest one dominates the MPS behavior approaches the $\Lambda$CDM model. 
For the case that one of the fields has the axion potential, the characteristic bump appeared when it had a light mass with large $\lambda_{\phi}$ values. On the contrary, if the mass is heavier, the bump disappeared regardless of the values of $\lambda_{\phi}$ and $R$. On the other hand, in the CMB power spectrum, we have not found significant changes unless one of the fields has $m_{\phi i} \leq 10^{-26}$ eV. That is, regardless of the number of fields we can always find a combination of the fields that matches the CMB observations where the  total contribution of dark matter is $\Omega_{\rm DM} = 0.264$, given by Planck 18 data \cite{Aghanim:2018eyx}.

We performed the parameter inference analysis with the Monte Python code \cite{Brinckmann:2018cvx, montepython} using BAO, Big Bang Nucleosynthesis, Ly-$\alpha$ forest and Supernovae for a single scalar field with the three potentials mentioned above and for double scalar fields. In the latter case, for simplicity we used the following combination of potentials: $V(\phi_1) = 1/2 m_{\phi 1}^2 \phi_1^2$ with $V(\phi_2) = 1/2 m_{\phi 2}^2 \phi_2^2$, $V(\phi_1) = 1/2 m_{\phi 1}^2 \phi_1^2$ with $V(\phi_2) = m_{\phi 2}^2f^2\left[1+\cos(\phi_2/f)\right]$ and $V(\phi_1) = m_{\phi 1}^2f^2\left[1+\cos(\phi_1/f)\right]$ with $V(\phi_2) = m_{\phi 2}^2f^2\left[1+\cos(\phi_2/f)\right]$. For the single case we found a bound for the mass values that corresponds to the one reported by the Ly-$\alpha$ data. We also presented the constrictions for the potential $V(\phi) = m_{\phi}^2f^2\left[\cosh(\phi/f)-1\right]$ where we have found a bound for $\lambda_{\phi}$ and, therefore, for the scalar field mass where, the latter, is also in agreement with the reported by the Ly-$\alpha$. In this case, the excluded values correspond to the cases in which the scalar field contributes to radiation as can be seen in Figure \ref{SFDM_cosh_Omegas}. For the double field models we found bounds for the masses and that the contribution of the fields depends on each other. Within our confidence regions are the masses reported in \cite{Broadhurst:2018fei}, in which more than one scalar fields is needed to explain the observed galaxies halos.

Finally, we have found that adding more scalar fields in order to explain astrophysical phenomena does not affect the known cosmology. So the MSFDM can be an alternative candidate to dark matter that is able to explain the observations at the cosmological and astrophysical levels. Its difference with other models is seen in the MPS at small scales. The results presented here can be generalized to larger number of fields with different potentials. We expect that forthcoming observations of collaborations such as DESI and LSST will allow us to better constrain the parameters of our model.

\acknowledgments

The authors thank Luis Ure\~na for the long and productive discussions about the paper. 
L.O.T.T acknowledge financial support from CONACyT doctoral fellowship. 
J.A.V. acknowledges the support provided by FOSEC SEP-CONACYT Investigaci\'on B\'asica A1-S-21925 and UNAM-DGAPA-PAPIIT IA104221.
This work was partially supported by CONACyT M\'exico under grants  A1-S-8742, 304001, 376127, 240512;
The authors are gratefully for the computing time granted by LANCAD and CONACYT in the
Supercomputer Hybrid Cluster "Xiuhcoatl" at GENERAL COORDINATION OF INFORMATION AND
COMMUNICATIONS TECHNOLOGIES (CGSTIC) of CINVESTAV.
URL: https://clusterhibrido.cinvestav.mx/ and Abacus clusters at Cinvestav, IPN;
I0101/131/07 C-234/07 of the Instituto
Avanzado de Cosmolog\'ia (IAC) collaboration (http://www.iac.edu.mx/).

\appendix
\section{Comparison with axionCAMB}
\begin{figure*}[t!]
	\centering
	 \makebox[\textwidth][c]{
	\includegraphics[trim = 0mm  0mm 10mm 0mm, clip, width=10cm, height=7.5cm]{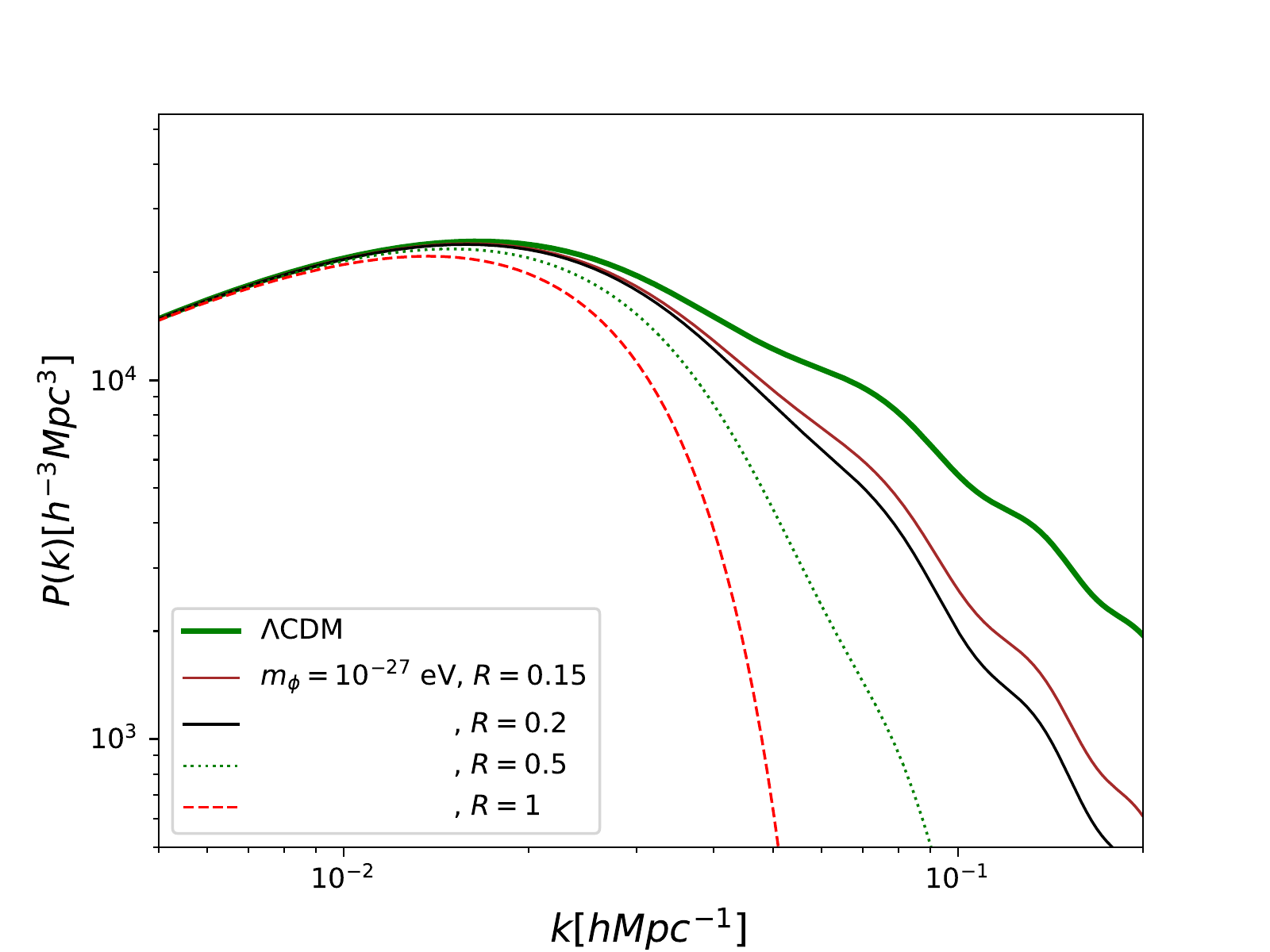}}
	\caption{\footnotesize{Evolution of the linear matter power spectrum at $z=0$ for a CDM+SFDM model with $V(\phi) = \frac{1}{2}m_{\phi}^2\phi^2$ with mass value $m_{\phi} = 10^{-27}$ eV and $R = 0.15$, $0.2$, $0.5$ and $1$. Colors and line types were chosen to match Figure 2 of \cite{Hlozek:2014lca} for an easy comparison.}} 
	\label{fig:app_SFDM_CDM}
\end{figure*}

In addition to CLASS, there is a code that solves the cosmology for an ultra-light axion, this is a modification to the CAMB code \cite{Lewis:1999bs, camb} called axionCAMB \cite{Hlozek:2014lca, axioncamb}. For the case where only the scalar field is present, the formalism presented in this work and the results of axionCAMB \cite{Hlozek:2014lca} are in agreement as mentioned in \cite{Urena-Lopez:2015gur}. Similarly, in \cite{Hlozek:2014lca}, the authors also studied the combination of an ultra-light axion with CDM. In Figure \ref{fig:app_SFDM_CDM}, we have reproduced the right side of figure 2 of the same reference for $R>0.1$ and $m_{\phi} = 10^{-27}$ eV. We found that the code presented here is in good agreement with the results of \cite{Hlozek:2014lca} except for small values of $R$, which translates to very small contributions from the scalar field ($\Omega_{\phi}\sim <0.02$).

\section{Extensions to MSFDM model: neutrino masses and curvature}
\begin{figure*}[t!]
	\centering
	 \makebox[\textwidth][c]{
	\includegraphics[trim = 0mm  0mm 10mm 0mm, clip, width=10cm, height=7.5cm]{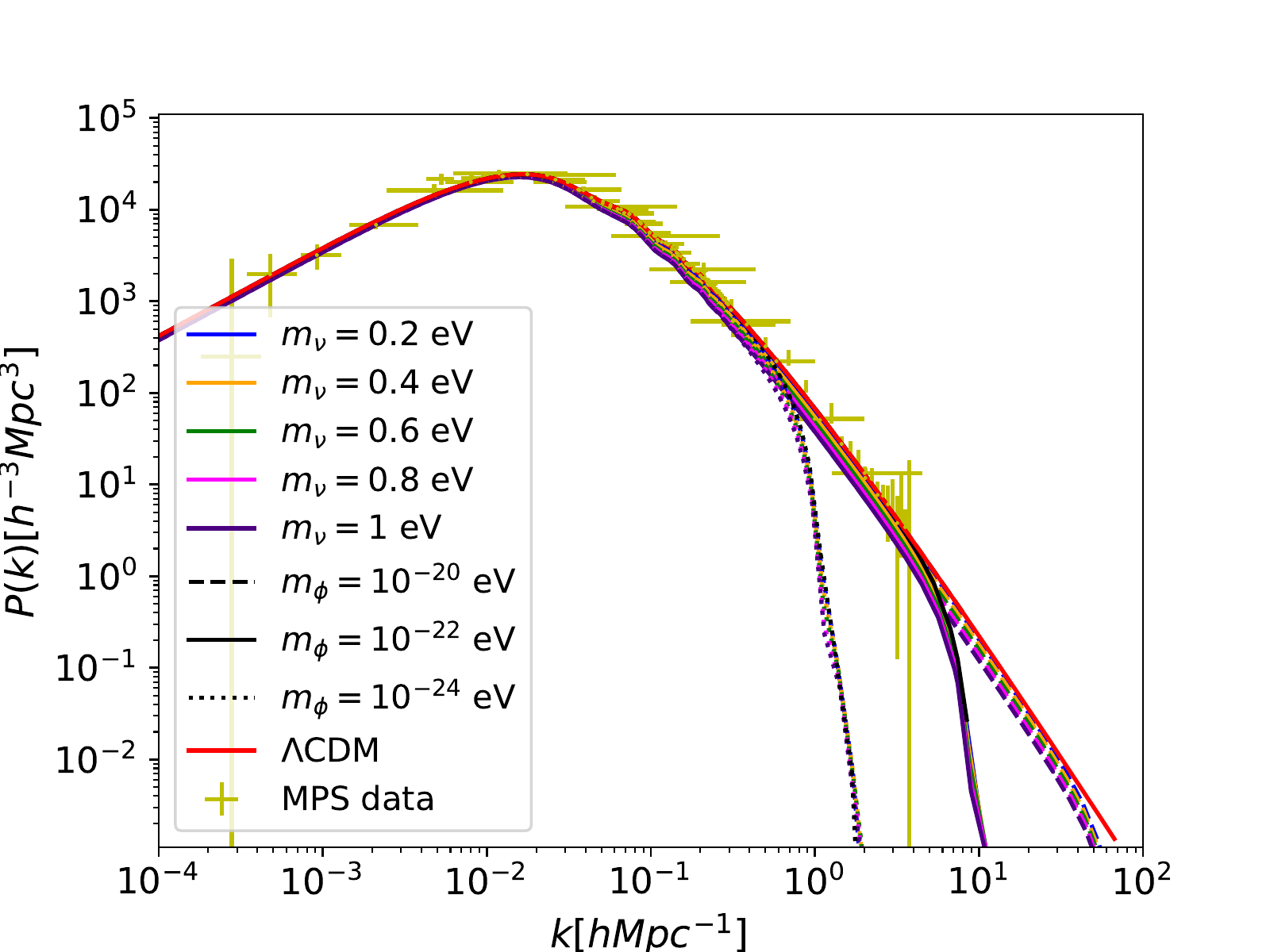}}
	\caption{\footnotesize{Evolution of the linear matter power spectrum at $z=0$ for $V(\phi) = \frac{1}{2}m_{\phi}^2\phi^2$ with neutrino masses $m_{\nu} = 0.2$, $0.4$, $0.6$, $0.8$ and $1$ eV. Coloured dotted lines represent the combinations with $m_{\phi} = 10^{-24}$ eV, coloured solid lines with $m_{\phi} = 10^{-22}$ eV and coloured dashed lines with $m_{\phi} = 10^{-20}$ eV. For example, the blue dashed line represents a model with $m_{\nu} = 0.2$ eV and $m_{\phi} = 10^{-20}$ eV. The black lines represent the case for the SFDM with masses $m_{\phi} = 10^{-24}$ eV (dotted), $m_{\phi} = 10^{-22}$ eV (solid) and $m_{\phi} = 10^{-20}$ eV (dashed) using the base model for neutrinos used in \cite{Aghanim:2018eyx}. Solid red line represents the MPS for $\Lambda$CDM. The data points for the MPS are the same used in section \ref{numerical} and were obtained from~\cite{Chabanier:2019eai}.}} 
	\label{fig:MPS_nu_SFDM}
\end{figure*}

\begin{figure*}[t!]
	\centering
	 \makebox[15cm][c]{
	\includegraphics[trim = 0mm  0mm 0mm 0mm, clip, width=10.5cm, height=7.5cm]{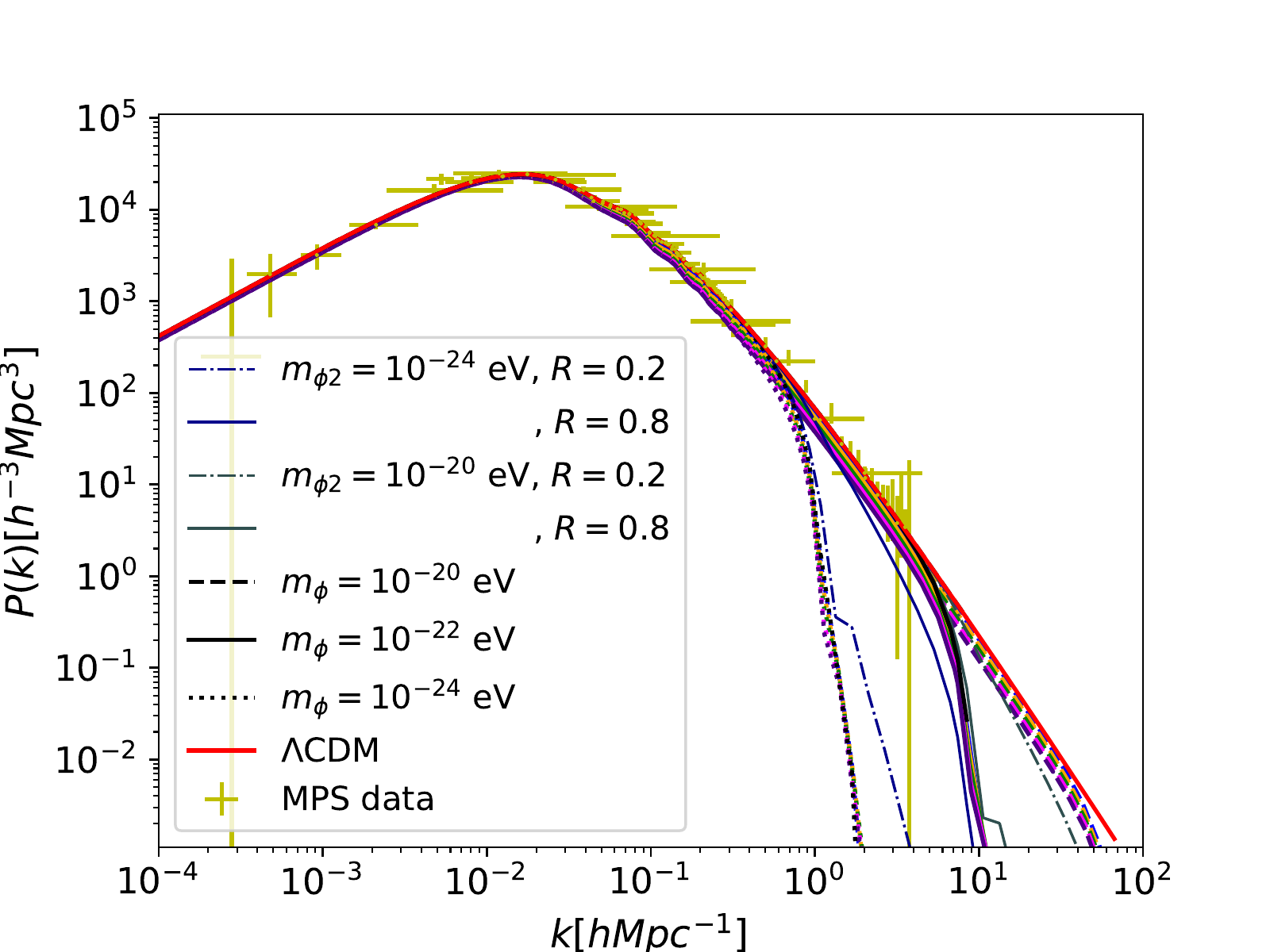}
	\includegraphics[trim = 0mm  0mm 0mm 0mm, clip, width=10.5cm, height=7.5cm]{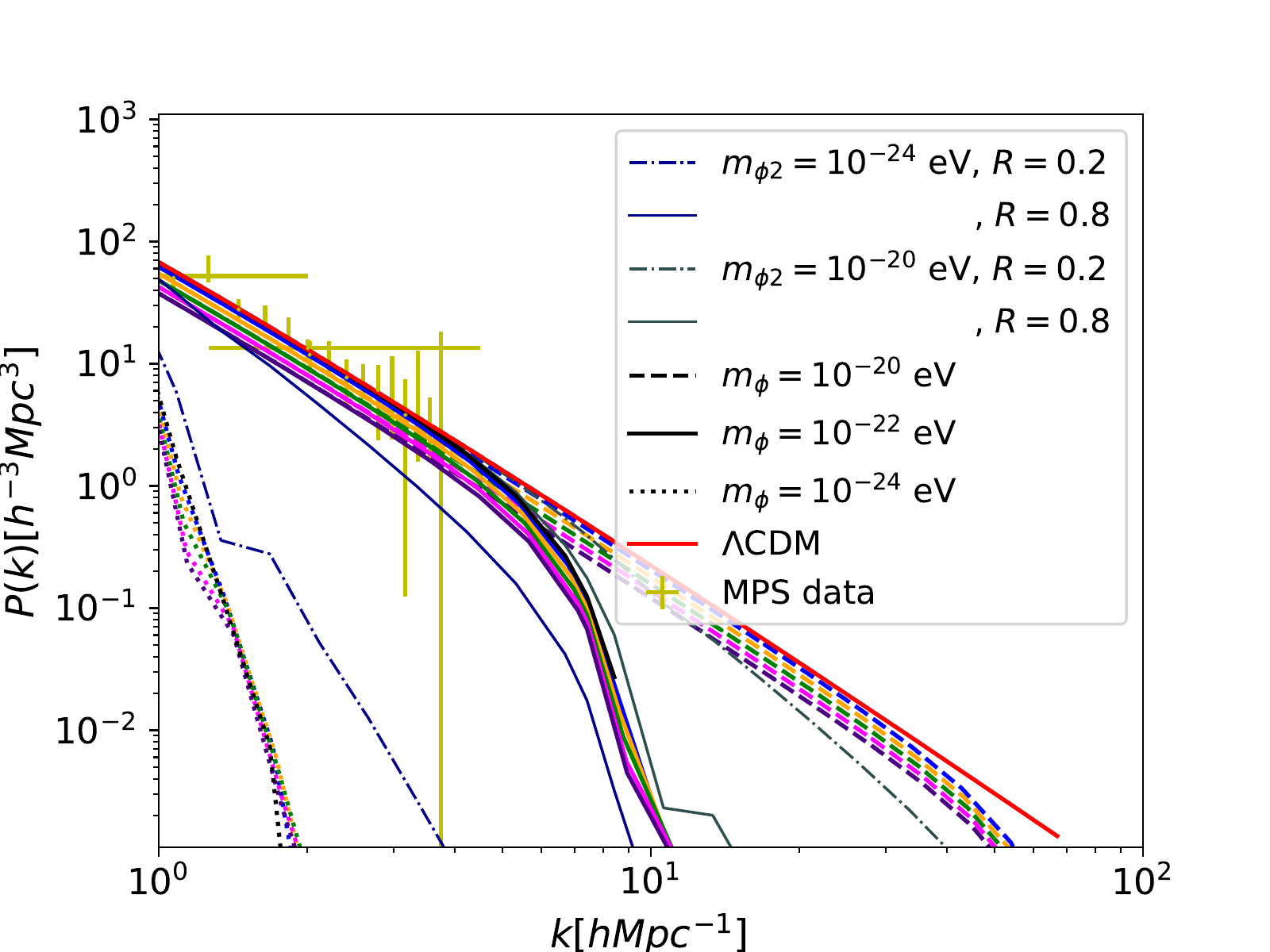}}
	
	\caption{\footnotesize{Evolution of the linear matter power spectrum at $z=0$ for a single field with potential $V(\phi) = \frac{1}{2}m_{\phi}^2\phi^2$ with neutrino masses $m_{\nu} = 0.2$ (blue), $0.4$ (orange), $0.6$ (green), $0.8$ (magenta) and $1$ eV (indigo) compared with the double field model with $V(\phi_{1,2}) = \frac{1}{2}m_{\phi 1,2}^2\phi_{1,2}^2$ using $m_{\phi 1} = 10^{-22}$ eV and different values of $m_{\phi 2}$ and $R$. For the single case varying neutrino mass, coloured dotted lines represent the combinations with $m_{\phi} = 10^{-24}$ eV, coloured solid lines with $m_{\phi} = 10^{-22}$ eV and coloured dashed lines with $m_{\phi} = 10^{-20}$ eV. The dark blue and gray lines represent the MSFDM evolution and solid red line represents the MPS for $\Lambda$CDM. The data points for the MPS are the same used in section \ref{numerical} and were obtained from~\cite{Chabanier:2019eai}. On the right side, a small-scale zoom is shown where you can see the consequence of changing the values of the neutrino masses and the scalar fields.}} 
	\label{fig:MPS_twoSFDM_vs_nu_SFDM}
\end{figure*}

It is well known that neutrinos have effects on the evolution of the Universe \cite{Lesgourgues:2006nd, Wong:2011ip}, in particular on the MPS \cite{2011MNRAS.410.1647A} where the MSFDM model has differences respect to $\Lambda$CDM. Therefore, in this section we vary the neutrino masses to see the effects on our model. We assume the neutrino model already presented in section \ref{sec:back-dyn}. In Figure \ref{fig:MPS_nu_SFDM} we show the evolution of the linear matter power spectrum for the SFDM model (one field) with the potential $V(\phi) = \frac{1}{2}m_{\phi}^2\phi^2$ with mass values $m_{\phi} = 10^{-24}$, $10^{-22}$ and $10^{-20}$ eV each combined with neutrino masses $m_{\nu} = 0.2$, $0.4$, $0.6$, $0.8$ and $1$ eV; bigger values are already discarded \cite{Liu:2018dsw, KATRIN:2019yun, Bayer:2021iyb, KATRIN:2021uub}. We found no significant differences respect to the neutrino base model. However, there could be degeneracies with the MSFDM. For example, in Figure \ref{fig:MPS_twoSFDM_vs_nu_SFDM} we compare the evolution of the MPS for the double field model, with potentials $V(\phi_{1,2}) = \frac{1}{2}m_{\phi 1,2}^2\phi_{1,2}^2$, and the single field case with different neutrino masses showed in Figure \ref{fig:MPS_nu_SFDM}. For the double field model, we fix the mass value of field one to $m_{\phi 1} = 10^{-22}$ eV while the second mass takes $m_{\phi 2} = 10^{-24}$ eV and $10^{-20}$ eV, with $R = 0.2$ and $R = 0.8$. We can see that the MPS of both models are very similar.

In the same way, we study the case where we add curvature to the scalar field model and find similar results as for $\Lambda$CDM. This can be seen in the left panel of Figure \ref{fig:app_log_mass} where we show the 1D posterior for $\Omega_k$. 

\section{MSFDM: triangle plots}
\label{ap:triangle}
\begin{figure*}[t!]
	\centering
	 \makebox[\textwidth][c]{
	\includegraphics[width=20cm]{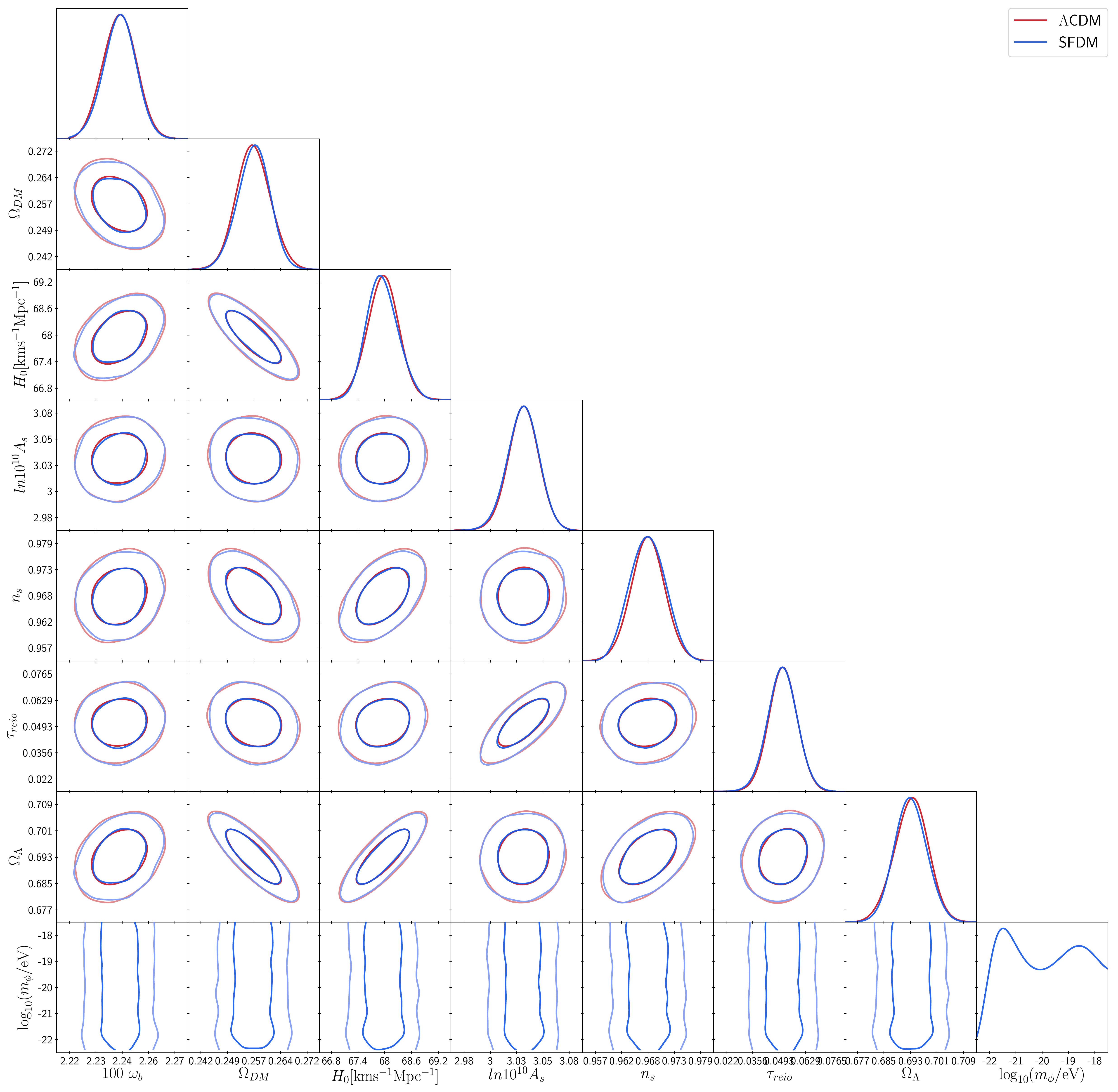}}
	\caption{\footnotesize{Triangle plots for the SFDM with potential $V(\phi) = \frac{1}{2}m_{\phi}^2\phi^2$ compared with $\Lambda$CDM as reference using the data from Planck 18, MPS from Ly-$\alpha$ BAO and SNe Ia. See text for details.}} 
	\label{fig:LCDM_SFDM}
\end{figure*}

\begin{figure*}[t!]
	\centering
	 \makebox[\textwidth][c]{
	 \includegraphics[width=5.5cm]{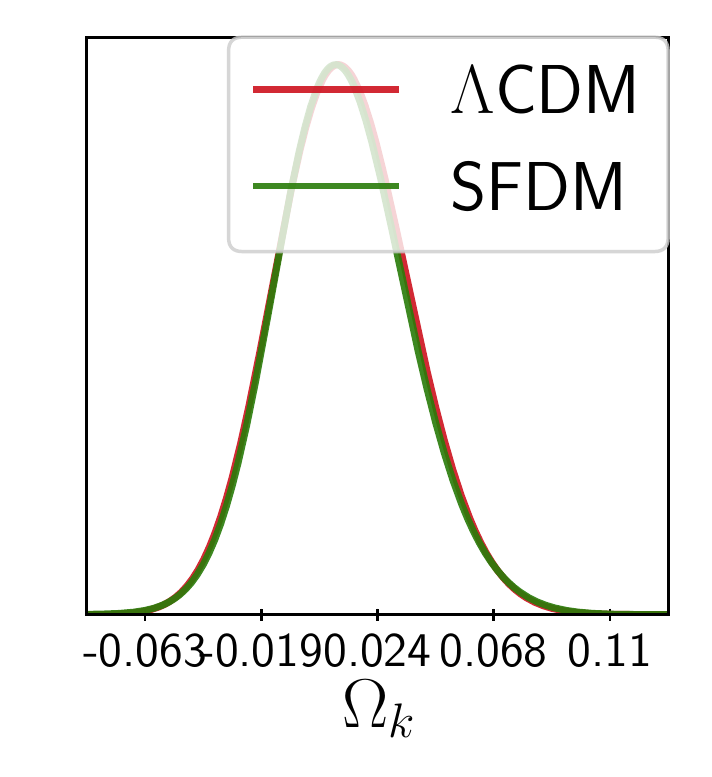}
	\includegraphics[width=5cm]{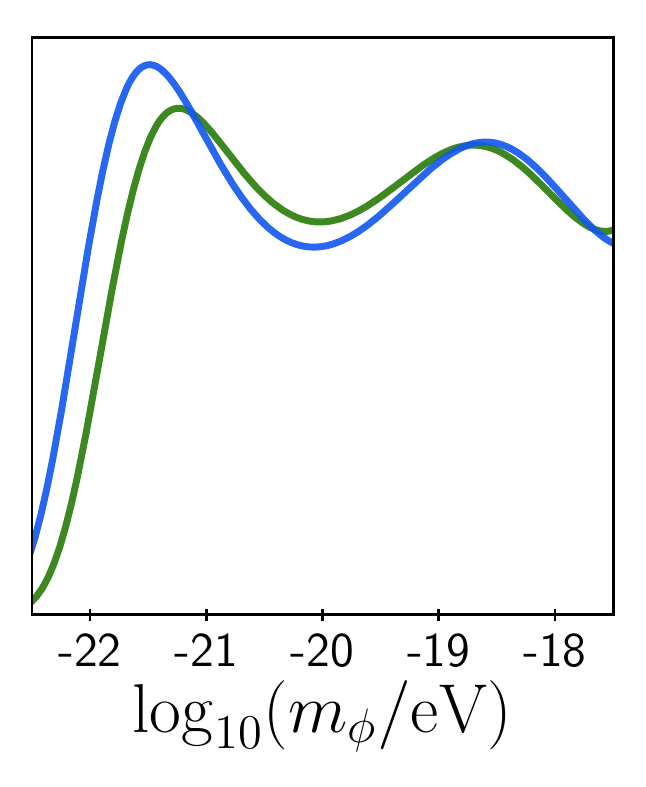}}
	\caption{\footnotesize{In the left panel we show the constraints on the curvature parameter for $\Lambda$CDM and for the SFDM model. In the right panel, we show the mass value constriction for the SFDM model with potential $V(\phi) = \frac{1}{2}m_{\phi}^2\phi^2$ using the data from Planck 18, MPS from Ly-$\alpha$, BAO and SNe Ia (blue) and, MPS from Ly-$\alpha$, BAO and SNe Ia (green).}} 
	\label{fig:app_log_mass}
\end{figure*}

In this appendix we show the triangle plots for the different potential combinations analyzed in section \ref{numerical}. In Figure \ref{fig:LCDM_SFDM} we show the posteriors of the physical baryon density parameter $\omega_{b,0}$, the logarithmic power spectrum scalar amplitude $\log\left(10^{10}A_s\right)$, the scalar spectral index $n_s$, the Thomson scattering optical depth due to reionization $\tau_{reio}$, the Hubble constant $H_0$ in $\mathrm{km}$ $\mathrm{s^{-1} Mpc^{-1}}$, the scalar field mass $\log_{10}\left(m_{\phi i}/ \mathrm{eV}\right)$, the Dark Energy density parameter $\Omega_{\Lambda}$ and the Dark Matter density parameter $\Omega_{DM}$. The last one refers to Cold Dark Matter for $\Lambda$CDM and to the scalar field for SFDM using Planck 18 data \cite{Aghanim:2018eyx, Planck:2019nip}, the 3D matter power spectrum inferred from Lyman-$\alpha$ data from BOSS and eBOSS collaboration \cite{Chabanier:2019eai}, the Ly-$\alpha$ BAO from eBOSS DR14 \cite{Cuceu:2019for}, the Galaxy BAO from DR12 \cite{Alam:2016hwk}, 6dFGS \cite{6dFGS} and SDSS DR7 \cite{Ross:2014qpa}, and the SNe Ia  survey Pantheon \cite{Scolnic:2017caz}. As we can see, there are no differences between the constrictions of the $\Lambda$CDM and SFDM parameters and they agree with the values reported in \cite{Aghanim:2018eyx}. Thus, in section \ref{cosmological} we decided to vary only the parameters corresponding to the scalar fields, leaving the rest of the parameters fixed. Using the fact that the SFDM model provides  similar constraints for the basic parameters using the Planck data and given that we found similar restrictions for the scalar field mass with and without Planck 18 data, we decided to use only the MPS from Ly-$\alpha$, BAO and SNe Ia data. This change in the data set used also gives us a reduction in computational time.

\begin{figure*}[t!]
	\centering
	\includegraphics[trim = 0mm  0mm 0mm 0mm, clip,width=12cm]{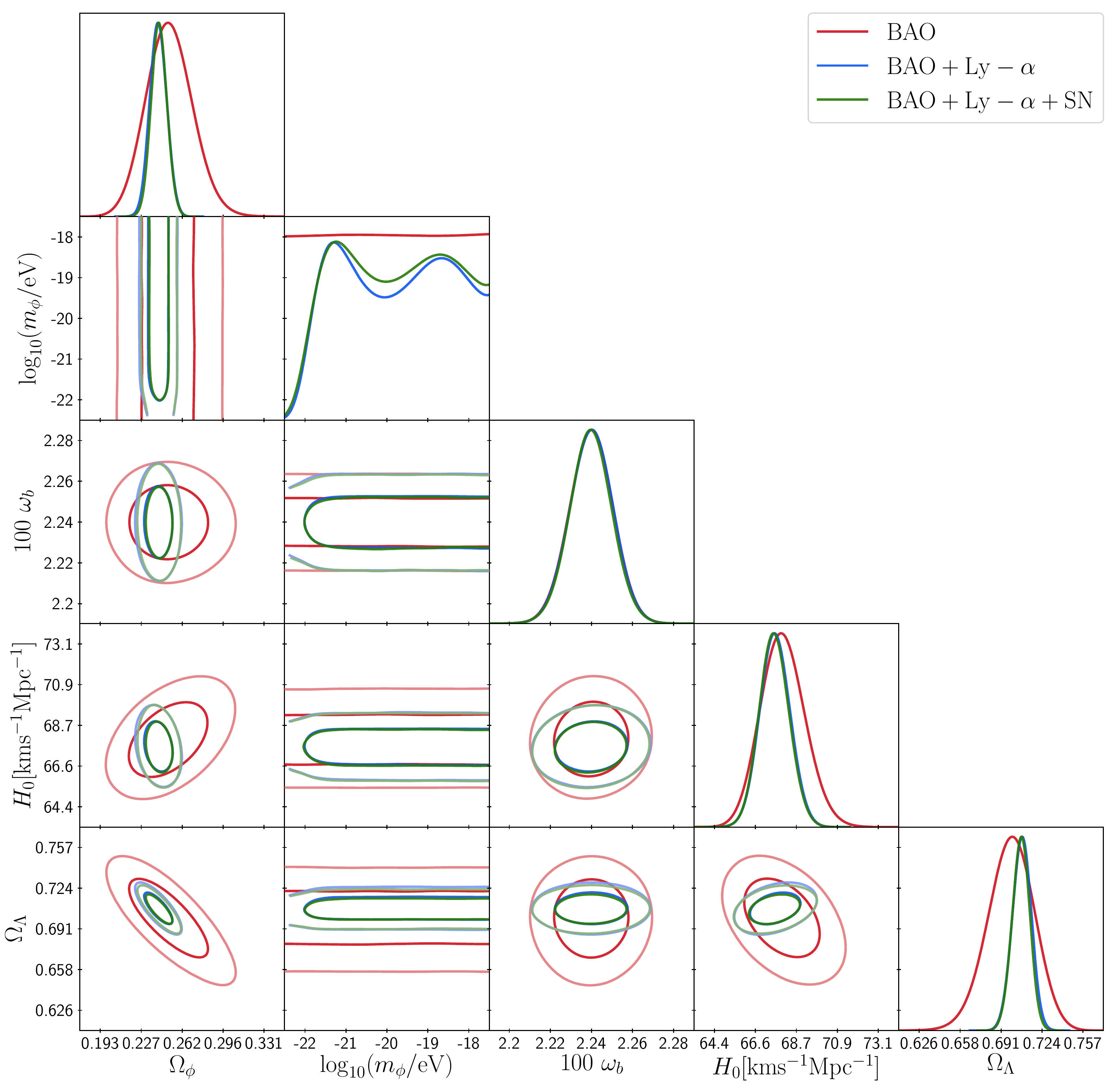}
	\caption{\footnotesize{1D and 2D posterior distribution for the potential $V(\phi) = \frac{1}{2}m_{\phi}^2\phi^2$. We found constrictions for $m_{\phi}$ and $\Omega_{\phi}$. We see that it is necessary to use different data sets to constrain the parameters of the scalar field. Here and in the following figures the Ly-$\alpha$ label refers to the MPS inferred from these data. See the text for details.}} 
	\label{fig:SFDM_quad_triangle}
\end{figure*}

\begin{figure*}[t!]
	\centering
	\includegraphics[trim = 0mm  0mm 0mm 0mm, clip, width=15cm]{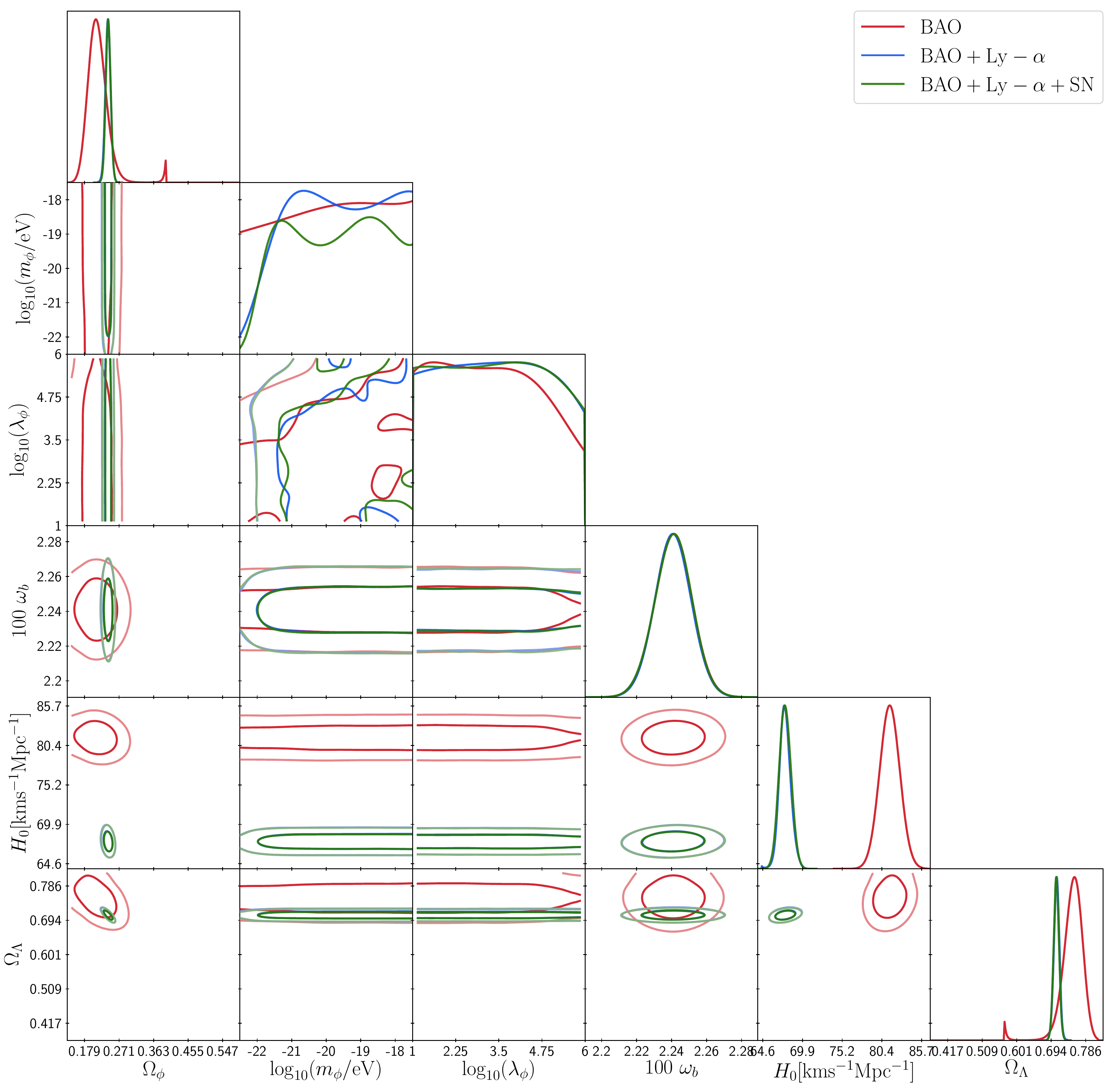}
	\caption{\footnotesize{1D and 2D posterior distribution for the potential $V(\phi) = m_{\phi}^2f^2\left[1+\cos(\phi/f)\right]$. We found constrictions only for $m_{\phi}$ and $\Omega_{\phi}$. It is necessary to use different data sets to constrain the parameters of the scalar field.}} 
	\label{fig:SFDM_cos_triangle}
\end{figure*}

\begin{figure*}[t!]
	\centering
	\includegraphics[trim = 0mm  0mm 0mm 0mm, clip, width=15cm]{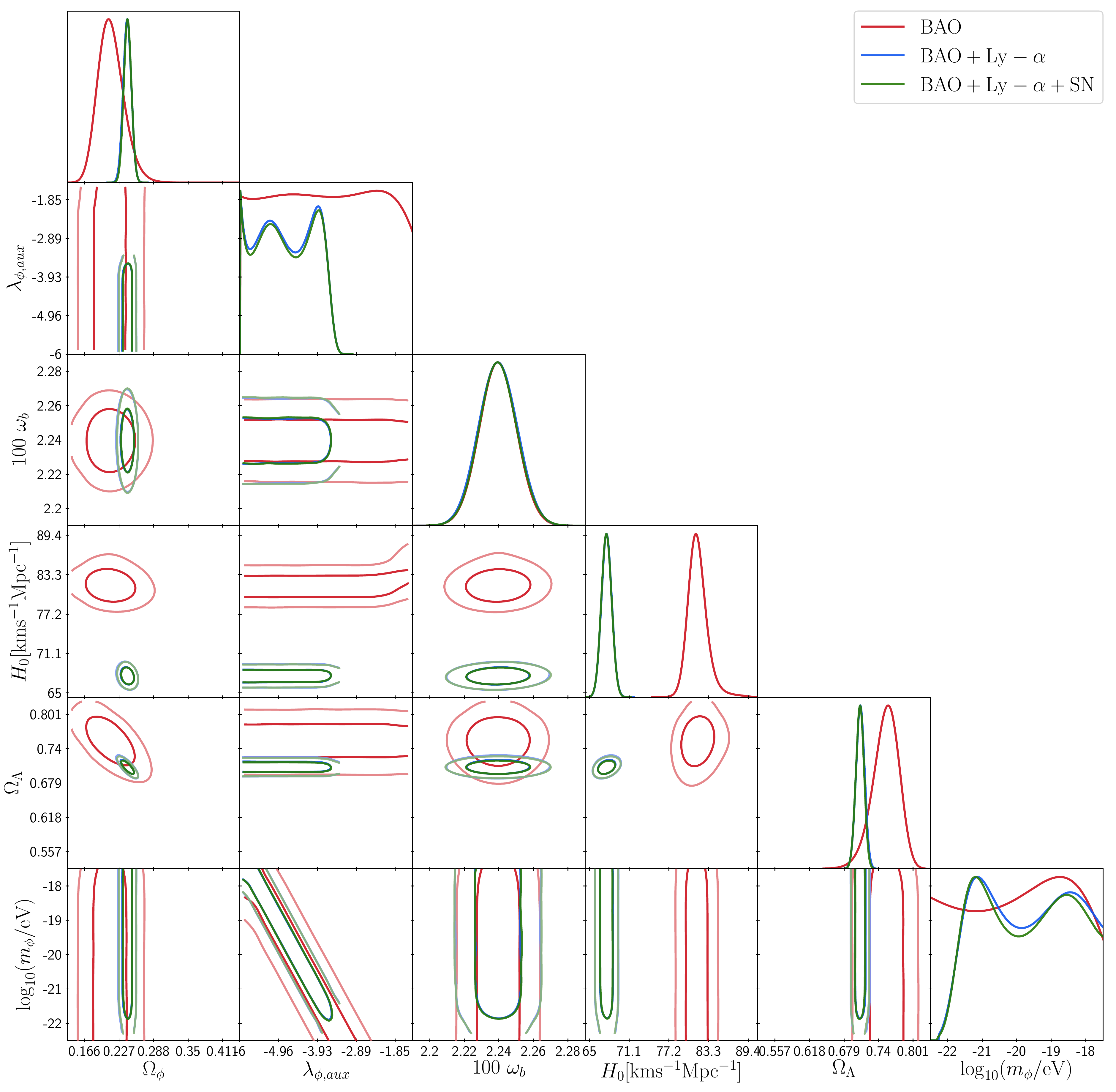}
	\caption{\footnotesize{1D and 2D posterior distribution for $V(\phi) = m_{\phi}^2f^2\left[\cosh(\phi/f)-1\right]$.}} 
	\label{fig:SFDM_cosh_triangle}
\end{figure*}

\begin{figure*}[t!]
	\centering
	\includegraphics[trim = 0mm  0mm 0mm 0mm, clip, width=17cm]{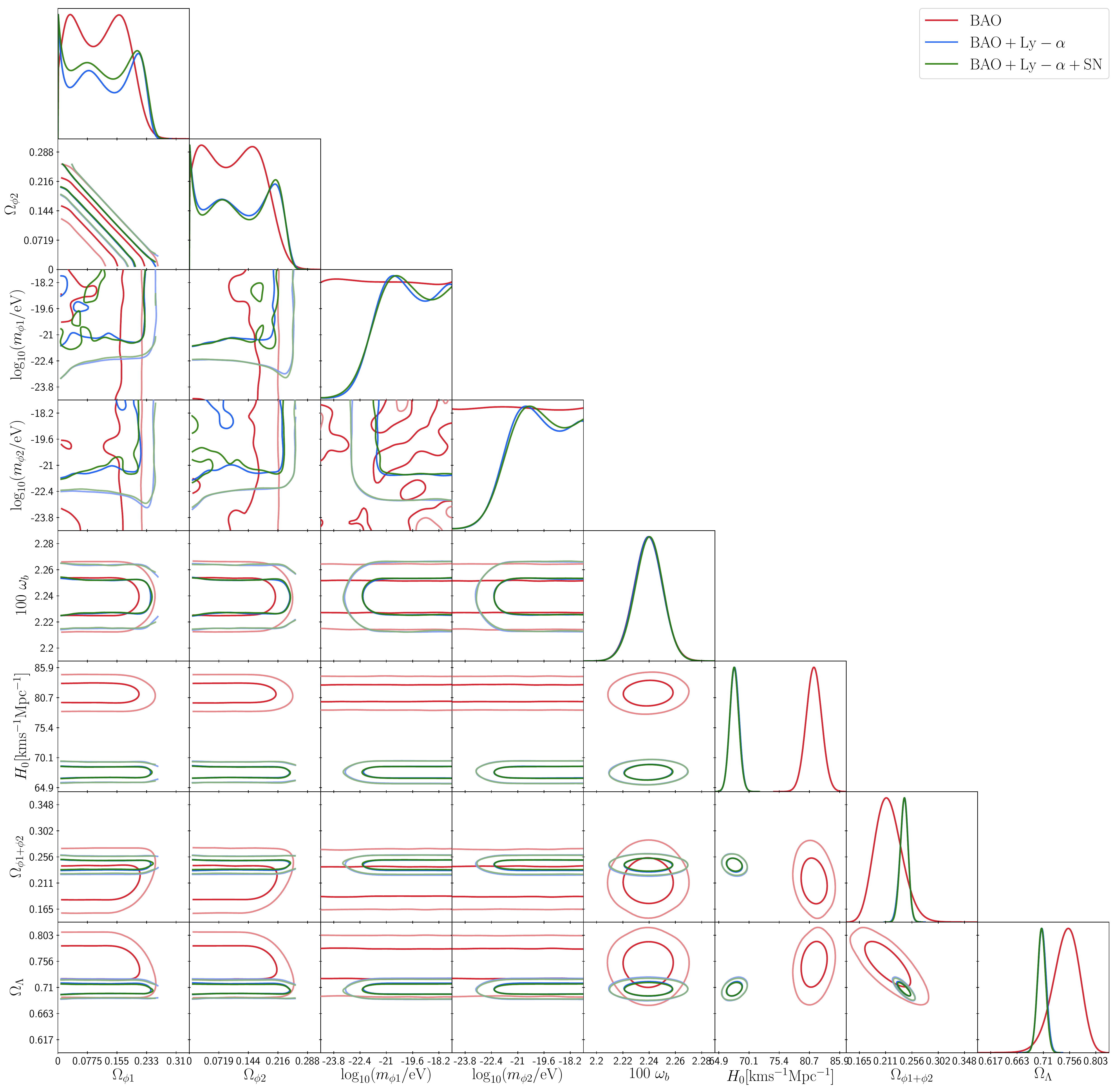}
	\caption{\footnotesize{1D and 2D posterior distribution of the fields considering potential $V(\phi_1) = 1/2 m_{\phi 1}^2 \phi_1^2$ for the first field with $V(\phi_2) = 1/2 m_{\phi 2}^2 \phi_2^2$}.} 
	\label{fig:SFDM_quad_quad_triangle}
\end{figure*}

\begin{figure*}[t!]
	\centering
	\includegraphics[trim = 0mm  0mm 0mm 0mm, clip, width=17cm]{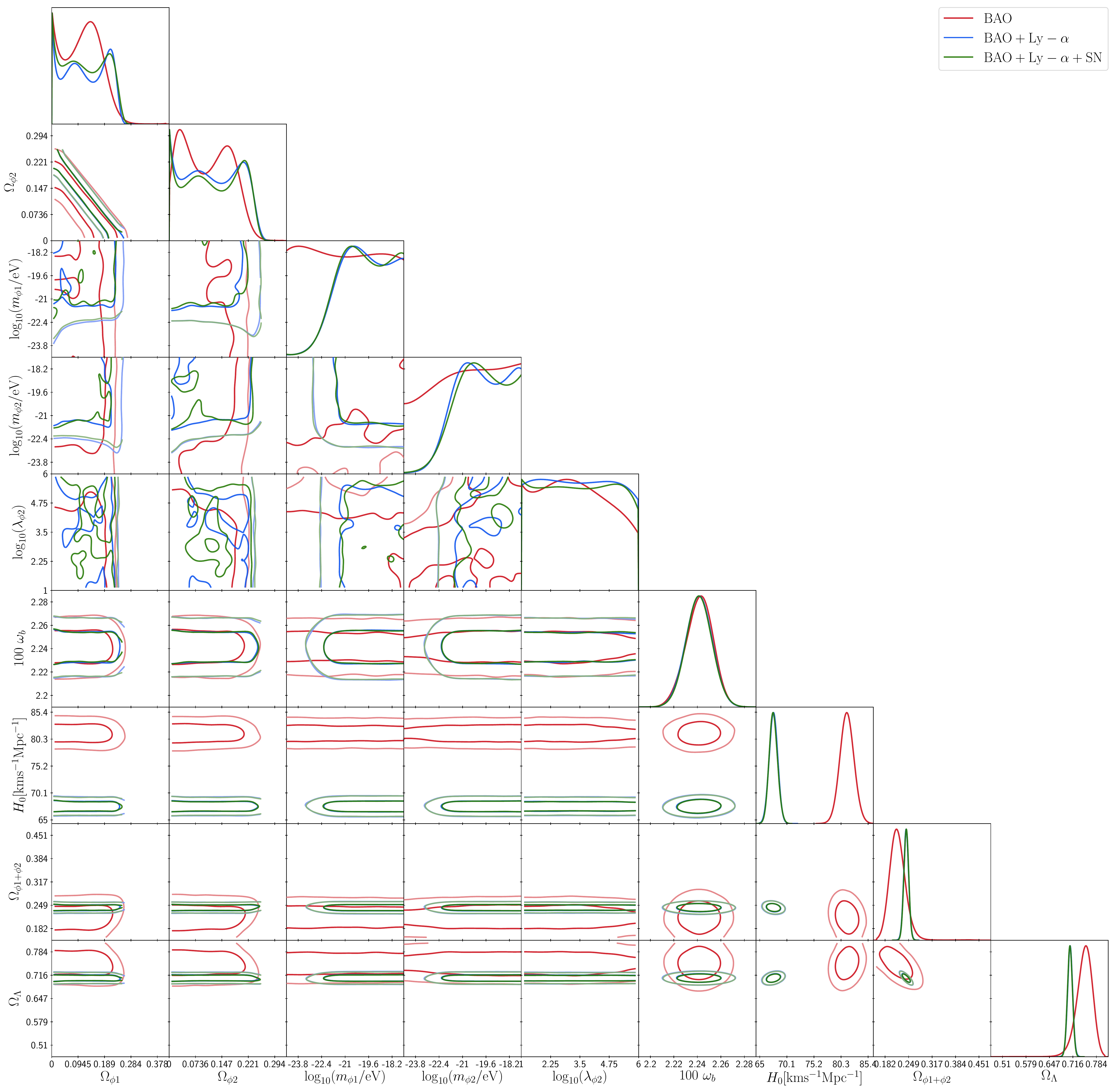}
	\caption{\footnotesize{1D and 2D posterior distribution of the fields considering potential $V(\phi_1) = 1/2 m_{\phi 1}^2 \phi_1^2$ for the first field with $V(\phi_2) = m_{\phi 2}^2f^2\left[1+\cos(\phi_2/f)\right]$.}} 
	\label{fig:SFDM_quad_cos_triangle}
\end{figure*}

\begin{figure*}[t!]
	\centering
	 \makebox[\textwidth][c]{
	\includegraphics[trim = 0mm  0mm 0mm 0mm, clip, width=17cm]{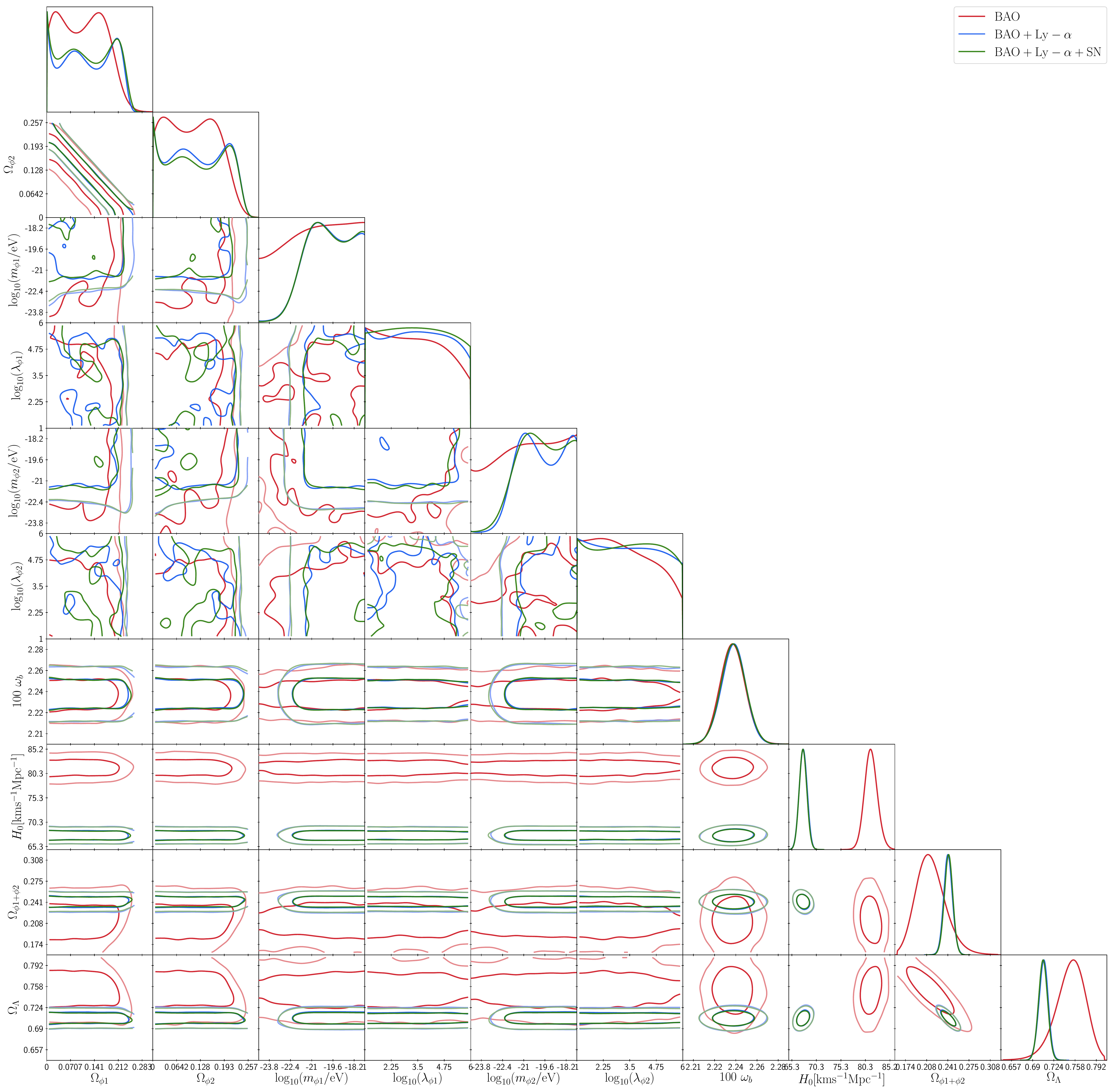}}
	\caption{\footnotesize{1D and 2D posterior distribution of the fields when both have the potential $V(\phi) = m_{\phi}^2f^2\left[1+\cos(\phi/f)\right]$.}} 
	\label{fig:SFDM_cos_cos_triangle}
\end{figure*}

On the other hand, in Figures \ref{fig:SFDM_quad_triangle}, \ref{fig:SFDM_cos_triangle}, \ref{fig:SFDM_cosh_triangle}, \ref{fig:SFDM_quad_quad_triangle}, \ref{fig:SFDM_quad_cos_triangle} and \ref{fig:SFDM_cos_cos_triangle} we also show the 1D and 2D posteriors of the combinations studied in section \ref{cosmological} when we consider each data set separately. In all cases we find that we need to combine the data sets to improve the constraints on the parameters of the scalar fields. Note that by using only the BAO data, the restriction for $H_0$ gives high values compared to Planck 18, so it is necessary to use more data sets in order to constrain this parameter. See \cite{Cuceu:2019for}.



\section{A general parametrisation}
\label{ap:abgparametrisation}
\begin{figure*}[t!]
	\centering
	 \makebox[\textwidth][c]{
	\includegraphics[trim = 0mm  0mm 10mm 0mm, clip, width=10cm, height=7.5cm]{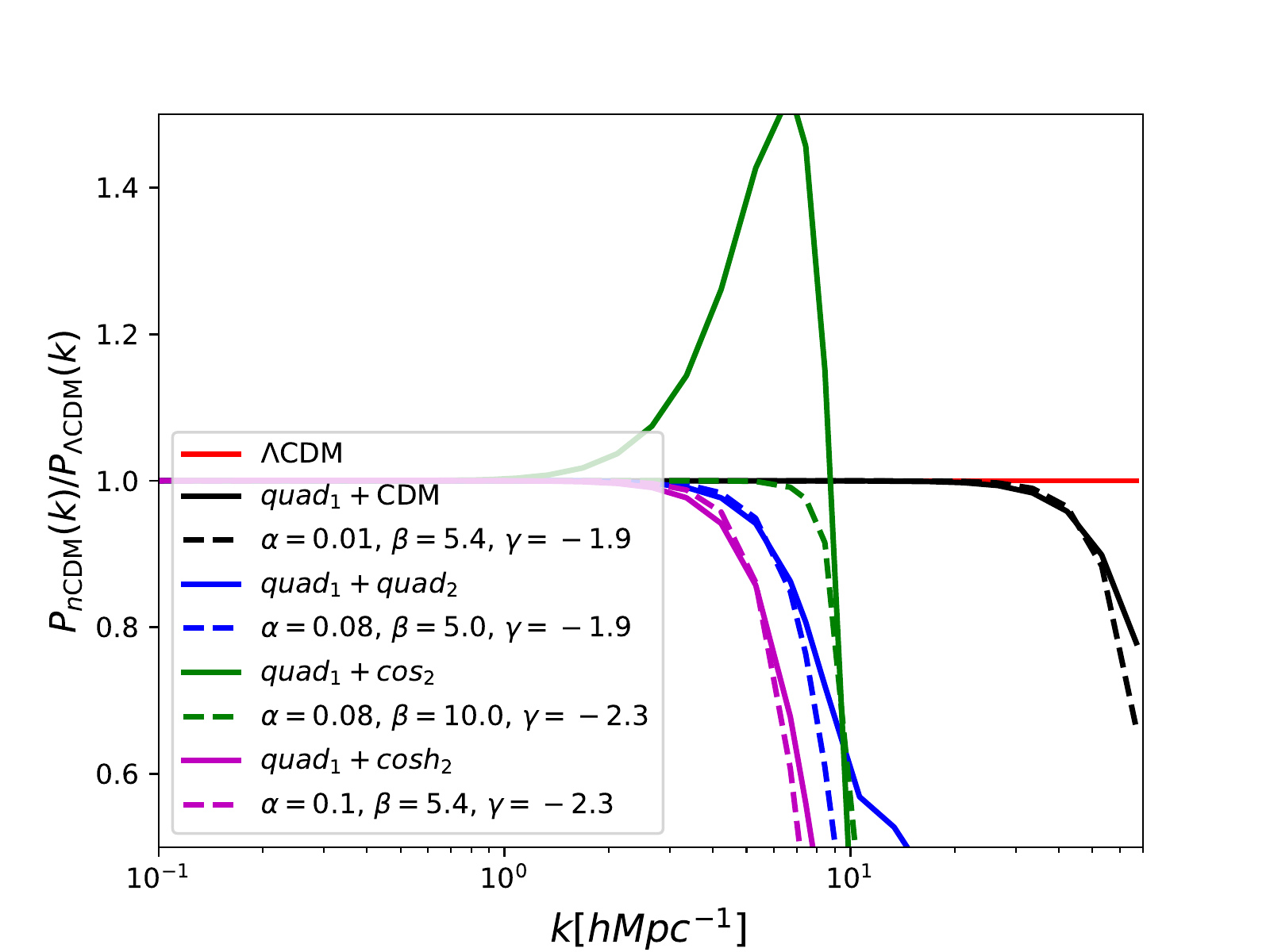}}
	\caption{\footnotesize{Transfer functions for the combinations SFDM+CDM with $m_{\phi}=10^{-20}$ eV and $R=0.2$ (black solid line); $quad_1 + quad_2$ with $m_{\phi_1}=10^{-22}$ eV, $m_{\phi 2}=10^{-20}$ eV and $R=0.2$ (blue solid line); $quad_1 + cos_2$ with $m_{\phi_1}=10^{-22}$ eV, $m_{\phi 2}=10^{-22}$ eV, $\lambda_{\phi 2}=10^5$ and $R=0.2$ (green solid line); and $quad_1 + cosh_2$ with $m_{\phi_1}=10^{-22}$ eV, $m_{\phi 2} = 0.3\times10^{-18}$ eV, $\lambda_{\phi 2} = -4\times10^5$ and $R = 0.5$. The dashed lines correspond to the transfer function using the $\alpha,\beta,\gamma$-parametrisation for different combinations of $\alpha$, $\beta$ and $\gamma$.}} 
	\label{fig:app_abg_parametrisation}
\end{figure*}

\lo{In \cite{Murgia2017, Murgia:2018now} the authors proposed a general parametrisation, called $\alpha,\beta,\gamma$-parametrisation, to describe the transfer function of non-CDM (nCDM) models, in particular, the fuzzy dark matter model which in our case corresponds to a single SFDM with the quadratic potential. In their studies they found that models allowed correspond those that meet $m_{\phi} \gtrsim 10^{-22}$ eV. In a first test, we found this parametrisation could also describe the combination of two fields with different combinations of potentials that  with the possible exception of the cases in which the characteristic bump of the trigonometric cosine-type potential occurs. In Figure \ref{fig:app_abg_parametrisation}, we show some examples of the transfer function for the combinations SFDM+CDM, $quad_1 + quad_2$, $quad_1 + cos_2$ and $quad_1 + cosh_2$ (solid lines) and the transfer function with the $\alpha,\beta,\gamma$-parametrisation (dashed lines). Although the values of the parameters $\alpha$, $\beta$ and $\gamma$ were set by hand, we can see they resemble the numerical results  we obtained with the exception of the cosine potential, where the characteristic bump cannot be described by this parametrisation. However, further studies similar to the one done in \cite{Archidiacono:2019wdp} are necessary to determine the relationship between the properties of the scalar fields and the $\alpha$, $\beta$ and $\gamma$ parameters and to obtain the respective constraints.}

\bibliographystyle{JHEP}

\providecommand{\href}[2]{#2}\begingroup\raggedright\endgroup

\end{document}